\documentclass[longauth]{aa}
\usepackage[toc,page]{appendix}
\usepackage{graphicx}
\usepackage{natbib}

\bibpunct{(}{)}{;}{a}{,}{,}
\usepackage{txfonts}

\begin{document}

   \title{PDRs4All\\ IX. Sulfur elemental abundance in the Orion Bar}
 
 \author{%
  Asunción Fuente \inst{\ref{CSIC}} \and
  Evelyne Roueff \inst{\ref{SorbonneLERMA}} \and
  Franck Le Petit \inst{\ref{SorbonneLERMA}} \and
  Jacques Le Bourlot \inst{\ref{SorbonneLERMA},\ref{ParisCite}} \and
  Emeric Bron \inst{\ref{SorbonneLERMA}} \and
  Mark G. Wolfire \inst{\ref{Maryland}} \and 
  James F. Babb \inst{\ref{Harvard}} \and 
  Pei-Gen Yan \inst{\ref{Harvard}} \and 
  Takashi Onaka \inst{\ref{TokyoAstro}} \and
  John~H.~Black \inst{\ref{ChalmersSpace}} \and
  Ilane	Schroetter \inst{\ref{ToulouseIRAP}} \and
  Dries Van De Putte \inst{\ref{STScI}} \and
  Ameek Sidhu \inst{\ref{WOntarioPA}, \ref{WOntarioIESE}} \and
  Amélie Canin \inst{\ref{ToulouseIRAP}} \and
  Boris Trahin \inst{\ref{SaclayIAS}} \and
  Felipe Alarcón \inst{\ref{Michigan}} \and
  Ryan Chown \inst{\ref{WOntarioPA}, \ref{WOntarioIESE}} \and
  Olga Kannavou \inst{\ref{SaclayIAS}} \and
  Olivier Berné \inst{\ref{ToulouseIRAP}} \and
  Emilie Habart \inst{\ref{SaclayIAS}} \and
  Els Peeters \inst{\ref{WOntarioPA}, \ref{WOntarioIESE}, \ref{CarlSagan}} \and
  Javier R. Goicoechea \inst{\ref{MadridFisFundamental}} \and 
  Marion Zannese \inst{\ref{SaclayIAS}} \and
  Raphael Meshaka \inst{\ref{SaclayIAS}, \ref{SorbonneLERMA}} \and
  Yoko Okada \inst{\ref{Koln}} \and
  Markus Röllig \inst{\ref{FrankfurtPhys}, \ref{Goethe}} \and
  Romane Le Gal \inst{\ref{GrenobleIPAG}, \ref{IRAM}} \and
  Dinalva A. Sales \inst{\ref{RioGrande}} \and
  Maria Elisabetta Palumbo \inst{\ref{INAFcatania}} \and
  Giuseppe Antonio Baratta \inst{\ref{INAFcatania}} \and
  Suzanne C. Madden \inst{\ref{SaclayAIMCEACNRS}} \and
  Naslim Neelamkodan \inst{\ref{Emirates}} \and
  Ziwei E. Zhang \inst{\ref{Riken}} \and
  P.C. Stancil \inst{\ref{Georgia}}
}

\institute{%
  Centro de Astrobiolog\'{\i}a (CAB), CSIC-INTA, Ctra. de Torrej\'on Ajalvir, km 4, 28850, Torrej\'on de Ardoz, Spain
  \label{CSIC} \and
  LERMA, Observatoire de Paris, PSL Research University, CNRS, Sorbonne Universités, F-92190 Meudon, France
  \label{SorbonneLERMA} \and
  Astronomy Department, University of Maryland, College Park, MD 20742, USA
  \label{Maryland} \and
  Center for Astrophysics \textbar\ Harvard \& Smithsonian, MS 14, 60 Garden St., Cambridge, MA 02138
   \label{Harvard} \and
  Department of Astronomy, Graduate School of Science, The University of Tokyo, 7-3-1 Hongo, Bunkyo-ku, Tokyo 113-0033, Japan
  \label{TokyoAstro} \and
  Department of Space, Earth and Environment, Chalmers University of Technology, Onsala Space Observatory, SE-439 92 Onsala, Sweden
  \label{ChalmersSpace} \and
  Institut de Recherche en Astrophysique et Planétologie, Université Toulouse III - Paul Sabatier, CNRS, CNES, 9 Av. du colonel Roche, 31028 Toulouse Cedex 04, France
  \label{ToulouseIRAP}         \and 
  Space Telescope Science Institute, 3700 San Martin Drive, Baltimore, MD 21218, USA
  \label{STScI} \and
  Department of Physics \& Astronomy, The University of Western Ontario, London ON N6A 3K7, Canada
  \label{WOntarioPA}     \and 
  Institute for Earth and Space Exploration, The University of Western Ontario, London ON N6A 3K7, Canada
  \label{WOntarioIESE}          \and 
  Institut d'Astrophysique Spatiale, Université Paris-Saclay, CNRS,  Bâtiment 121, 91405 Orsay Cedex, France
  \label{SaclayIAS}        \and 
  Department of Astronomy, University of Michigan, 1085 South University Avenue, Ann Arbor, MI 48109, USA
  \label{Michigan} \and
  Carl Sagan Center, SETI Institute, 339 Bernardo Avenue, Suite 200, Mountain View, CA 94043, USA
  \label{CarlSagan}         \and
  Instituto de Física Fundamental  (CSIC),  Calle Serrano 121-123, 28006, Madrid, Spain
  \label{MadridFisFundamental} \and
  I. Physikalisches Institut der Universität zu Köln, Zülpicher Stra{\ss}e 77, 50937 Köln, Germany
  \label{Koln} \and
  Physikalischer Verein - Gesellschaft für Bildung und Wissenschaft, Robert-Mayer-Str. 2, 60325 Frankfurt, Germany
  \label{FrankfurtPhys} \and 
  Goethe-Universität, Physikalisches Institut, Frankfurt am Main, Germany
  \label{Goethe} \and
  Institut de Planétologie et d'Astrophysique de Grenoble (IPAG), Université Grenoble Alpes, CNRS, F-38000 Grenoble, France
  \label{GrenobleIPAG} \and
  Institut de Radioastronomie Millimétrique (IRAM), 300 Rue de la Piscine, F-38406 Saint-Martin d'Hères, France
  \label{IRAM} \and
  Instituto de Matemática, Estatística e Física, Universidade Federal do Rio Grande, 96201-900, Rio Grande, RS, Brazil
  \label{RioGrande} \and
  INAF - Osservatorio Astrofisico di Catania, Via Santa Sofia 78, 95123 Catania, Italy
  \label{INAFcatania} \and
  AIM, CEA, CNRS, Université Paris-Saclay, Université Paris Diderot, Sorbonne Paris Cité, 91191 Gif-sur-Yvette, France
  \label{SaclayAIMCEACNRS} \and
  Department of Physics, College of Science, United Arab Emirates University (UAEU), Al-Ain, 15551, UAE
  \label{Emirates} \and
  Star and Planet Formation Laboratory, RIKEN Cluster for Pioneering Research, Wako, Saitama 351-0198, Japan
  \label{Riken} \and
  Department of Physics and Astronomy, University of Georgia, Athens, GA 30602-2451, USA
  \label{Georgia} \and
  Universit\'e Paris Cit\'e \label{ParisCite}
  }

 \abstract 
   {One of the main problems in astrochemistry is determining the amount of sulfur in volatiles and refractories in the interstellar medium. The detection of the main sulfur reservoirs (icy H$_2$S and atomic gas) has been challenging, and estimates are based on the reliability of models to account for the abundances of species containing less than 1\% of the total sulfur. The high sensitivity of the \textit{James Webb} Space Telescope provides an unprecedented opportunity to estimate the sulfur abundance through the observation of 
the [S I] 25.249 $\mu$m  line.}
   {Our aim is to determine the amount of sulfur in the ionized and warm molecular phases toward the Orion Bar as a template to investigate sulfur depletion in the transition between the ionized gas and  the molecular cloud in HII regions. }
   {We used the [S III] 18.7 $\mu$m, [S IV] 10.5 $\mu$m,  and [S l] 25.249 $\mu$m lines to estimate the amount of sulfur in the ionized and molecular gas along the Orion Bar. For the theoretical part, we used an upgraded version of the Meudon photodissociation region (PDR) code to model the observations. New inelastic collision rates of neutral atomic sulfur with ortho- and para- molecular hydrogen were calculated to predict the line intensities. }
  {The [S III] 18.7 $\mu$m and [S IV] 10.5 $\mu$m lines are detected over the imaged region with a shallow increase (by a factor of 4) toward the HII region. This suggests that their emissions are partially coming from the Orion Veil. We estimate a moderate sulfur depletion, by a factor of $\sim$2, in the ionized gas.  The corrugated interface between the molecular and atomic phases gives rise to several edge-on dissociation fronts we refer to as DF1, DF2, and DF3. The [S l] 25.249 $\mu$m line is only detected toward DF2 and DF3, the dissociation fronts located farthest from the HII region. This is the first ever detection of the [S l] 25.249 $\mu$m line in a PDR. The detailed modeling of DF3 using the Meudon PDR code shows that the emission of the  [S l] 25.249 $\mu$m line  is coming from warm ($>$ 40 K) molecular gas located at A$_{\rm V}$ $\sim$ 1$-$5 mag from the ionization front. Moreover, the intensity of the  [S l] 25.249 $\mu$m line is only accounted for if we assume the presence of undepleted sulfur. }
 {Our data show that  sulfur remains undepleted along the ionic, atomic, and molecular gas in the Orion Bar. This is consistent with recent findings that suggest that sulfur depletion is low in massive star-forming regions because of the interaction of the UV photons coming from the newly formed stars with the interstellar matter. }
  
   \maketitle

\section{Introduction}
\label{intro}
Gas and solids experience a continuous evolution from their early times in molecular clouds until their
incorporation into a growing planet through a protoplanetary disk. Although this evolution lasts for a few million years, we now think that the disk's chemical composition is to a large extent determined by the physical and
chemical conditions in the progenitor molecular cloud (see, e.g., \citealp{Guzman2017, Jensen2019, Booth2021}). Elemental abundances are preserved during the
planet formation process, and their values in the progenitor protoplanetary disk will  determine the chemical composition of future
planet atmospheres. The origin of the raw material needed to facilitate life is
not clear, however. The Earth could have acquired its volatiles and organics from other sources. About 0.5\%
of the Earth’s mass could have come from the bombardment of cometary bodies from the outer Solar System
during the so-called Late Veneer \citep{Wang2013, Ehrenfreund2015, Caselli2020}. Since the composition of comets seems to be similar to that of the dust grains in
the initial protosolar nebula \citep{Capria2017, Bockelee2017}, we need to follow the whole history of elements in space, from
molecular clouds to protoplanetary disks, in order to understand how heavy atoms are incorporated into moons and planets
and eventually life-forms.

\begin{figure*}
\includegraphics[width=0.55\textwidth]{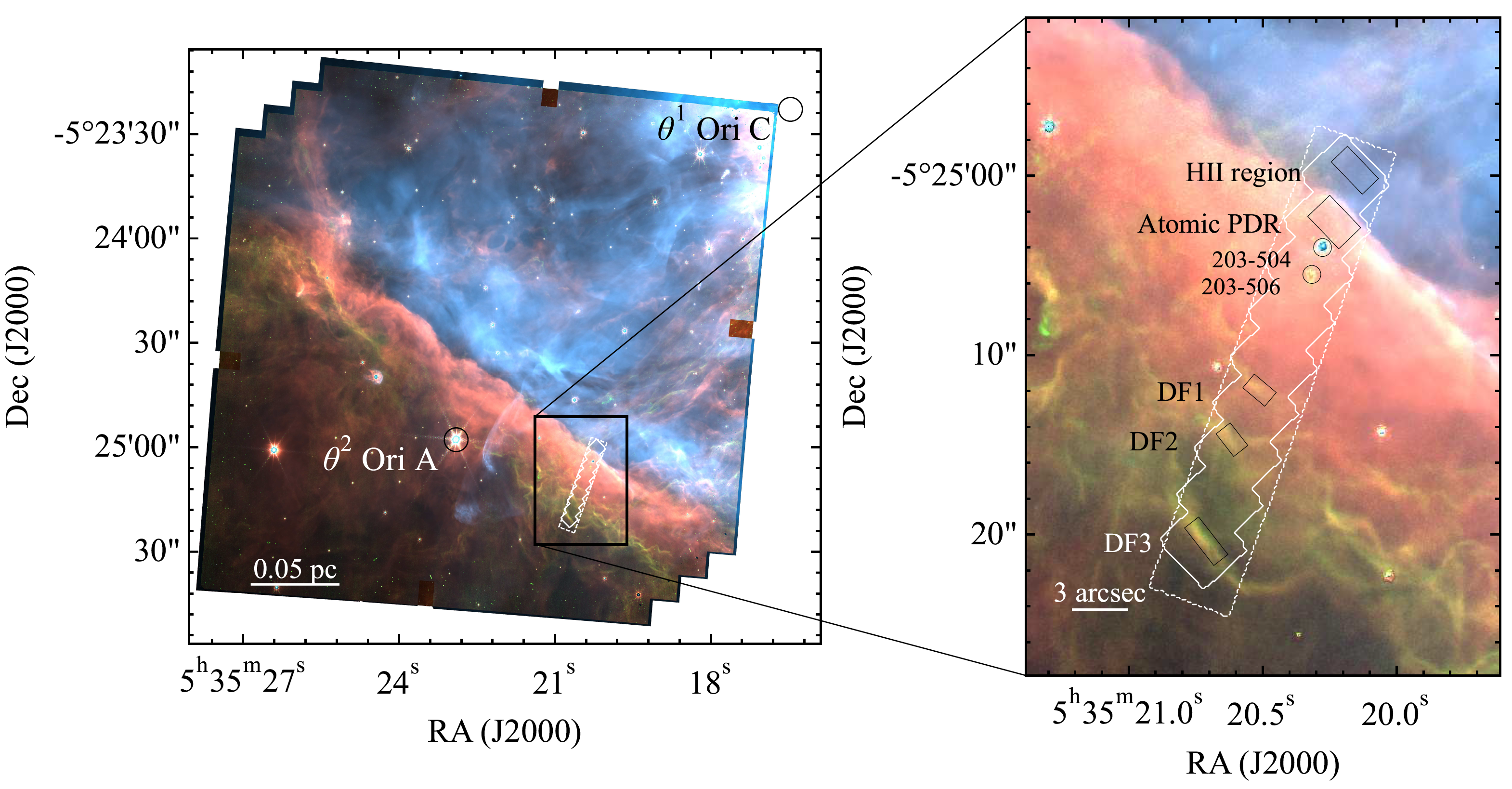}
\includegraphics[width=0.45\textwidth]{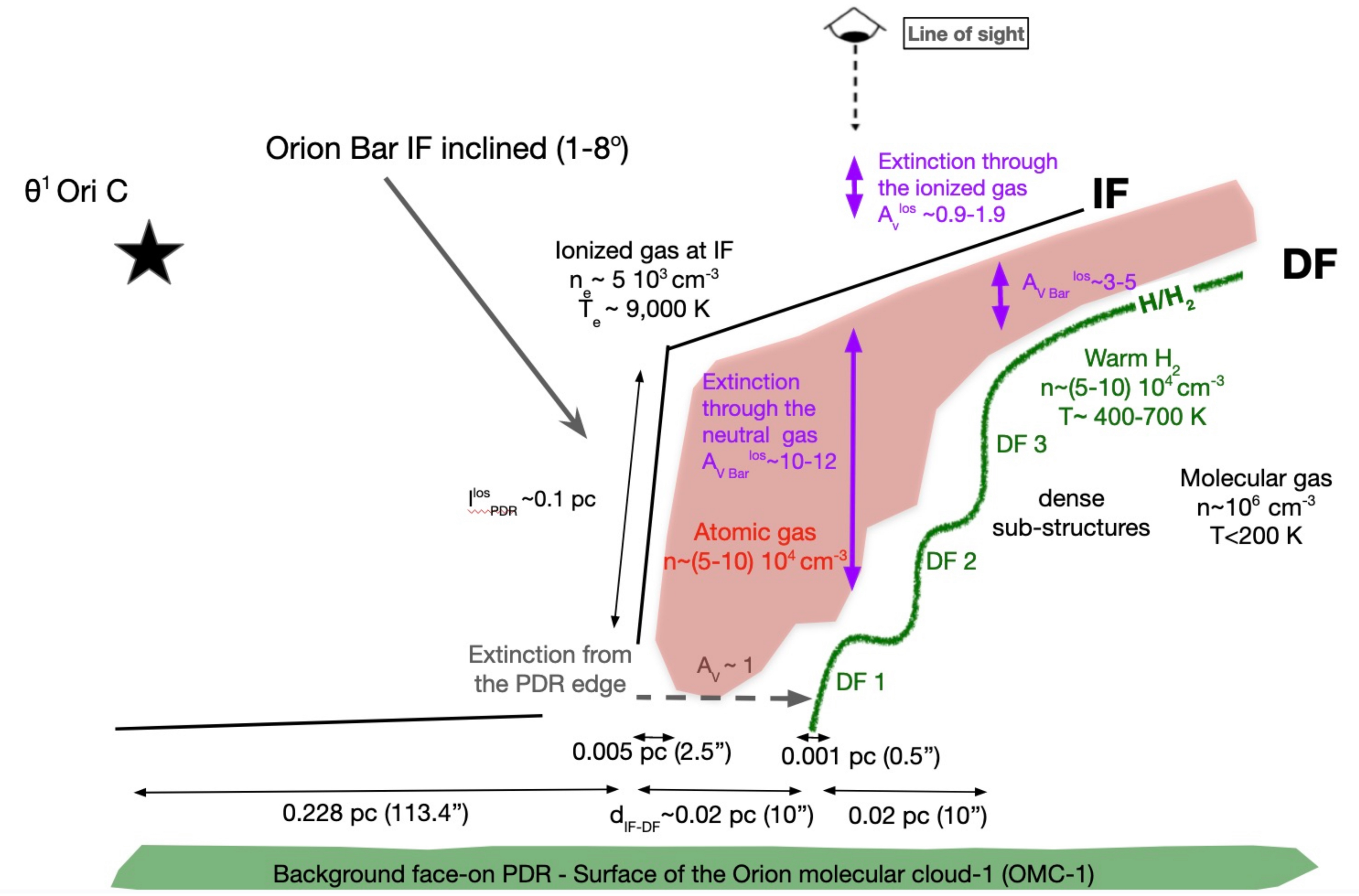}
\caption{Overview of the NIRSpec and MIRI MRS mosaics observed in PDRs4All and scheme of the geometry of the Orion Bar. Left panel: Composite NIRCam image encoded as follows: F335M (AIB emission) in red, F470N-F480M (H$_2$ emission) in green, and F187N (Paschen $\alpha$ emission) in blue \citep{Habart2023, Peeters2023}. The footprints of the NIRSpec and MIRI MRS mosaics are shown as solid and dashed lines, respectively. The positions of DF1, DF2, and DF3, as defined by \citet{Habart2023}, together with those of the protostars d203-504 and d203-506 are indicated. This figure has been taken from \citet{Chown2023}.  Right panel: Sketch of the Orion Bar adapted from \citet{Habart2023} and \citet{Peeters2023}.}
\label{fig1}
\end{figure*}

Sulfur is one of the most abundant elements in the Universe and is known to play a significant role in biological systems in Earth. In the interstellar medium, sulfur is found in gas phase, in interstellar ices, and in the refractory material that eventually can form terrestrial planets. The elemental abundances defined as
the fraction of a  given element in volatiles (gas + ice) are essential parameters for understanding the chemistry of interstellar gas and the composition of the refractory grain cores.
However, while the carbon and oxygen budgets in star-forming regions have been extensively studied, there are still  a lot of open questions 
regarding the amount of sulfur and its relative distribution between volatiles and refractories in the interstellar medium. Observations in diffuse clouds are consistent
with essentially undepleted sulfur \citep{Sofia1994, Neufeld2015}. Moderate sulfur depletions (by less than a factor of 10) are also consistent with observations of sulfur-bearing species in some photodissociation regions (PDRs; \citealt{Goicoechea2006, Goicoechea2021, GoicoecheaCuadrado2021}).
However, the sum of the observed abundances of sulfur molecules in gas and ice only constitutes $<$5\% of the expected amount in cold regions \citep{Vastel2018, Riviere2019, Fuente2019}, and sulfur depletion seems to be larger than a factor of $\sim$10 in the shielded and densest regions of dark clouds based on model analyses (e.g., L1544: \citealp{HilyBlant2022}; \citealp{Fuente2023}).
 The given evolutionary stage at which this big depletion occurs and the form in which sulfur is incorporated into (semi-)refractory material is still unknown. For instance, it has been suggested that the so-called depleted sulfur might be locked in refractory FeS 
\citep{Kama2019}. Sulfur could also form semi-refractory material such as sulfur allotropes \citep{JimenezEscobar2011, JimenezEscobar2012, Fuente2019, Shingledecker2020, Cazaux2022}. Different sulfur reservoirs would translate into different fractions of sulfur in volatiles in protoplanetary disks, which would lead to different compositions of planets and paths for sulfur leading to life. Understanding the details of sulfur depletion is therefore an essential question in astronomy.

Chemical models predict that 88\% of the sulfur in cold molecular clouds (n$_{\rm H}$ = 2$\times$10$^4$ cm$^{-3}$, T = 10 K)  is in atomic form at a typical age of 0.1 Myr  (see, e.g., \citealp{Vidal2017-DC}). However,  due to its high excitation conditions (see Table~\ref{lines}), the [S I] 25.249 $\mu$m line has only been detected in bipolar outflows \citep{Anderson2013} thus far, precluding accurate knowledge of its abundance in most environments and the full testing of our chemical models.
PDRs are regions where most of the gas is neutral (i.e., H or H$_2$) and their physical and chemical conditions are determined by far-ultraviolet (FUV) photons (i.e., 6 eV $<$ E $<$ 13.6 eV) emitted by massive stars. PDRs are found in essentially all relevant astrophysical environments, including  diffuse clouds, protoplanetary disks, molecular cloud surfaces, globules, planetary nebulae, and starburst galaxies. Investigating sulfur chemistry in PDRs is of paramount importance to disentangling the problem of missing sulfur since PDRs constitute the transition between the sulfur undepleted ionized gas and the heavily depleted molecular cold gas.

The \textit{James Webb} Space Telescope (JWST; \citealp{Gardner2006}) Early Release Science (ERS) program $``$PDRs4All: Radiative feedback from massive stars$"$ has observed the prototypical PDR usually referred to as the Orion Bar as a template for the study of Galactic and extragalactic PDRs \citep{Berne2022}.
The Orion  Bar is the sharp edge bordering the HII region M~42, which is located on the near side of the giant molecular cloud OMC~1, at a distance D~=~414$\pm$7 pc. \citep{Menten2007}.
Observations of this so-called Orion Bar in radio-continuum, vibrationally excited H$_2$ \citep{Hayashi1985} and at 3.3 $\mu$m \citep{Bregman1994} show that the bar is formed by the gas associated with an almost edge-on ionization front (IF). Because of this favorable geometry and its closeness to the Solar System, the Orion Bar is probably the best studied PDR and an ideal test bench for chemical models.
The gas associated with the Orion Bar has been widely studied at all wavelengths \citep{Omodaka1994, Tauber1994, Tauber1995, Hogerheijde1995, White1995, Fuente1996, Fuente2003}. \citet{Burton1990} modeled the Orion Bar with a density n$_{\rm H}$ = 10$^5$ cm$^{-3}$ and an UV intensity field of $\sim$ 1 $\times$ 10$^5$ in Habing field units using available near- and mid-infrared observations \citep{Hayashi1985, Peimbert1977, Haas1986}. Based on \textit{Herschel} data, \citet{Goicoechea2011} and \citet{Joblin2018} conclude that the emission of the OH and CO far-infrared lines can only originate from small structures with typical thicknesses of a few times 10$^{-3}$ pc and at high thermal pressures (P$_{\rm th}$ $\sim$10$^8$ K cm$^{-3}$). PDRs4All data have provided, for the first time, the possibility to resolve the IF and photodissociation fronts (DFs)  in the Orion Bar \citep{Habart2023, Peeters2023, Chown2023}. We now know that the Orion Bar is not a single edge-on PDR but a series of edge-on and face-on PDRs formed in the corrugated surface of the molecular cloud surrounding the HII region M~42 (see the sketch in Fig.~\ref{fig1}).

This paper aims to characterize the sulfur chemistry in the ionized and warm molecular gas associated with the Orion Bar to eventually determine the sulfur depletion along the border of the molecular cloud.  The [S IV]~10.5~$\mu$m and  [S III]~18.7~$\mu$m lines are used to trace the ionized gas, and a comparison of the H$_2$ S(1) and [S I]~25.249~$\mu$m lines allows us to determine sulfur depletion in the molecular phase. Combining these results with previous estimates of the sulfur abundance in the M~42 nebula using optical lines \citep{Esteban2004, Daflon2009} and the estimation in the S$^+$/S transition region using the sulfur recombination lines from \citet{GoicoecheaCuadrado2021}, we are able to determine the amount of sulfur in all the gaseous phases across the Orion Bar.  

\begin{table}
\caption{Studied MIRI lines in order
of increasing wavelength with atomic data from NIST~\citep{NIST_ASD}.  Line parameters of the H$_2$ line are from \citet{Roueff2019}}
\label{lines}
\centering
\begin{tabular}{ccccc}\\
\hline
$\lambda$ &  Species & \multicolumn{1}{c}{Transition} &    A$_{ij}$               &  E$_u$   \\
($\mu$m)  &          & \multicolumn{1}{c}{Lower - Upper}           &    (s$^{-1}$)   &   (K)  \\
\hline
10.5105   &  [S IV]             &  ${}^2P_{1/2}^{\textrm{o}}-{}^2P_{3/2}^{\textrm{o}}$       &    7.74  $\times$ 10$^{-3}$      & 1368.9       \\     
17.0348   &  H$_2$  v=0  &  J=1 - J=3                             &    4.76 $\times$ 10$^{-10}$       &  1015.1     \\                                               
18.7129   &  [S III]            & ${}^3P_1-{}^3P_2$               &    2.06  $\times$ 10$^{-3}$      &  1198.7     \\                                               
25.2490   &   [S I]             &   ${}^3P_2-{}^3P_1$             &   1.40  $\times$ 10$^{-3}$       &   569.83    \\                                                
\hline 
\end{tabular}
\end{table}

\begin{figure*}
\includegraphics[angle=0,scale=.6]{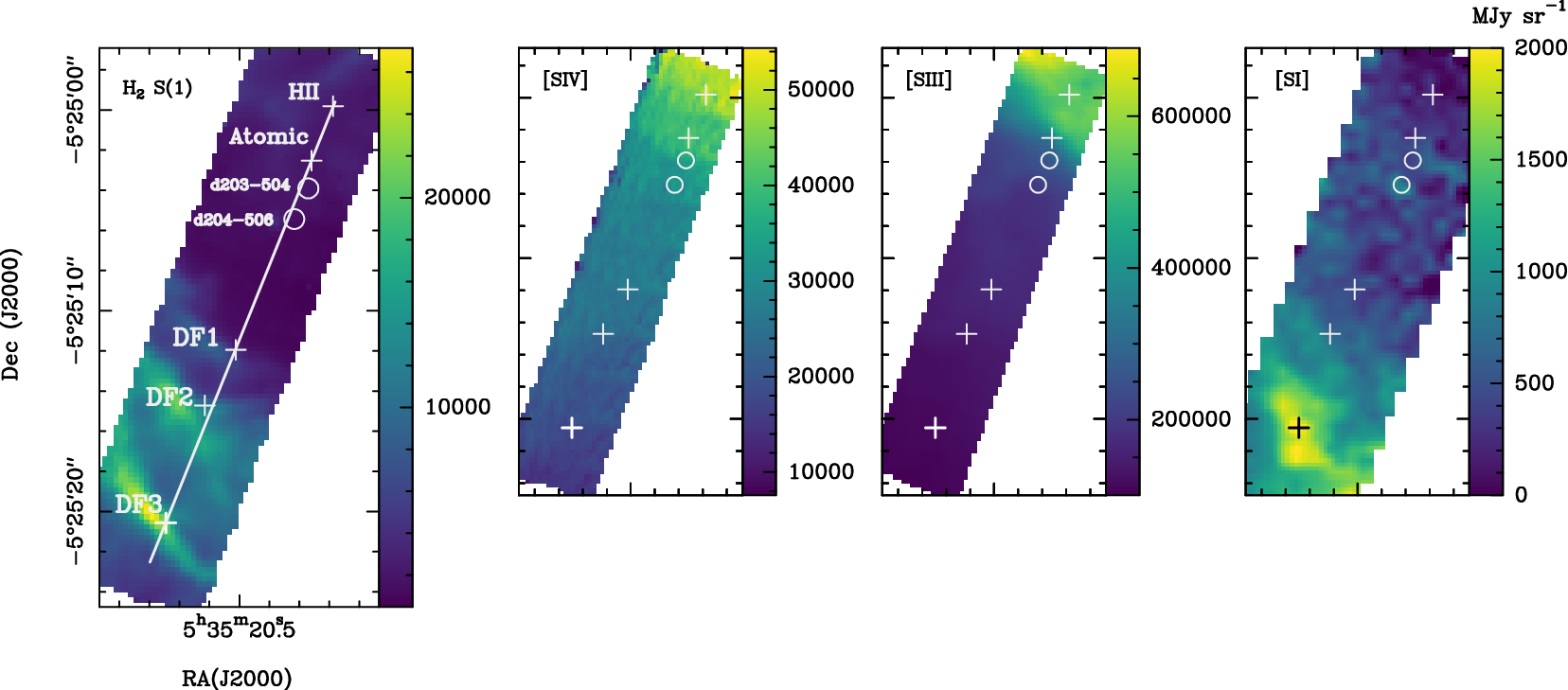}
\caption{Peak intensity maps of the lines indicated in the top-left corner of each panel. All the maps have been re-grided to a pixel of $\sim$0.3$"$.  Crosses indicate the positions of HII, Atomic, DF1, DF2, and DF3, as listed in Table~\ref{fields}. Circles indicate the positions of the protoplanetary disks d203-504 and d203-506 \citep{Bally2000}. The straight line indicates the cut shown in Fig.~\ref{cuts}. The rms in each map is: 80~MJy~sr$^{-1}$ (H$_2$ S(1)), 50~MJy~sr$^{-1}$ (~[SIV]~), 80~MJy~sr$^{-1}$ (~[S III]~), and 273~MJy~sr$^{-1}$ (~[S I]).}
\label{maps}
\end{figure*}

\begin{figure}
\includegraphics[angle=0,scale=.47]{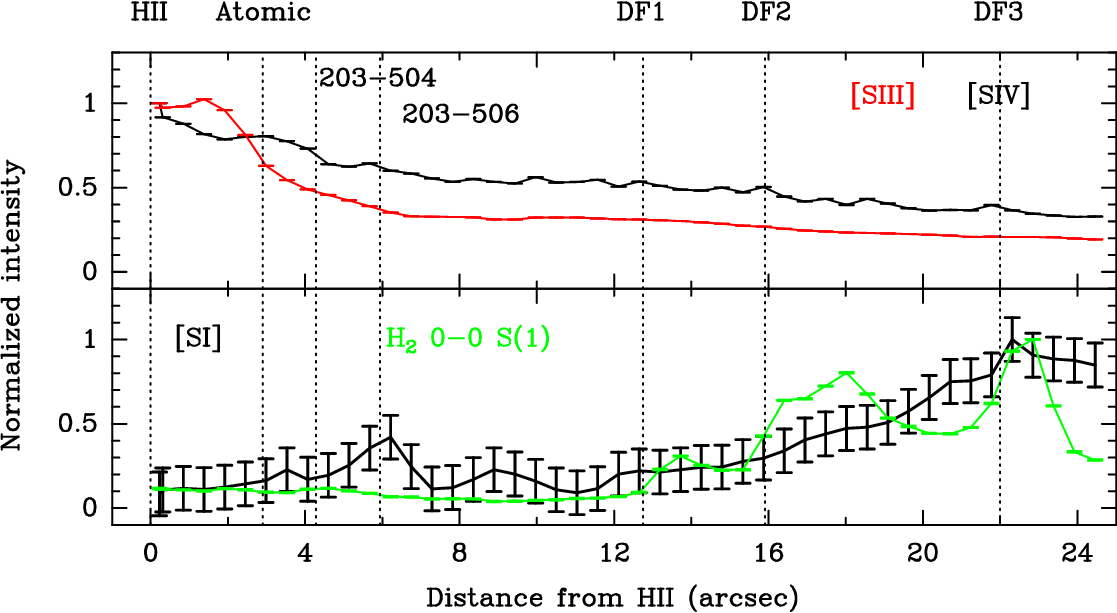}
\caption{Cuts along the straight line starting at HII and crossing all the DFs drawn in the H$_2$ S(1) panel of Fig.~\ref{maps}. In all cases, the intensity has been normalized to 1. 
Dashed lines indicate the positions of HII, Atomic, DF1, DF2, DF3, and the protostellar objects d203$-$504 and d203$-$506.}
\label{cuts}
\end{figure}

\section{Observations}
\label{obs}
This paper is based on observations of  the ERS program “PDRs4All: Radiative feedback from massive stars” (ID1288)\footnote{pdrs4all.org; https://www.stsci.edu/jwst/science-execution/program-information.html?id=1288} PIs: Bern\'e, Habart, Peeters; \citealp{Berne2022}). Within this program, the Orion Bar PDR has been observed
with NIRSpec and the Mid-Infrared Instrument (MIRI) in medium resolution spectroscopy (MRS) mode \citep{Wells2015, Argyriou2023}. MIRI observations were performed over a 1~$\times$9 pointing mosaic,  which defines a rectangle almost perpendicular to the Orion Bar (see Fig.~\ref{fig1} and \citealp{Chown2023}). This region was observed in all four MRS channels (channels 1, 2, 3, and 4), and all three sub-bands within each channel (short, medium, and long). 
Our resulting 3D maps cover the full MRS wavelength range (4.90 to 27.90 $\mu$m) with a spectral resolution ranging from R$\sim$3700 in channel 1 to $\sim$1700 in channel 4 and a spatial resolution of 0.207$"$ at short wavelengths to 0.803$"$ at long wavelengths, corresponding to 86 and 332 AU, respectively at the distance of the Orion Nebula.
This paper is focused on the study of the H$_2$ S(1) 17.035 $\mu$m,  [S IV] 10.5 $\mu$m, [S III] 18.7 $\mu$m,  and [S I] 25.249 $\mu$m lines, all placed in channel 4.

The MIRI/MRS observations were processed from uncalibrated data through the official pipeline, using 1.12.5 of the \texttt{jwst}  Python package, with the \texttt{jwst\_1147.pmap} Calibration Reference Data System (CRDS) context. We made additional corrections after the default pipeline to improve the quality of  our dataset. A detailed description of all these modifications can be found in \citet{Putte2024}. In current data cubes,  the wavelength calibration  is accurate up to a few km s$^{-1}$ at short wavelengths, and about 30 km s$^{-1}$ at the longest wavelength \citep{Argyriou2023}. The pointing accuracy is about 0.45\arcsec\ without target acquisition, and the typical accuracy of the assigned coordinate system is about 0.3\arcsec, where the main source of uncertainty is the guide star catalog \citep{Patapis2024}.
The spectrophotometric calibration was initially based on a single standard star \citep{Gordon2022}, but data from additional stars were recently introduced into the calibration.
The inclusion of these additional calibration data (since \texttt{jwst\_1094.pmap}), led to an improved matching of the continuum flux between the four channels of MIRI/MRS, with only minor flux offsets in the overlap regions between the channels \citep{Putte2024}.
Together with support for time-dependent calibrations, this postprocessing much improved the continuum flux and calibration of channel 4 where our lines are found. This improvement was specially important in the case of the  weaker [S I] 25.249 $\mu$m line.
As reported in the JWST user documentation\footnote{https://jwst-docs.stsci.edu/jwst-calibration-pipeline-caveats/jwst-miri-mrs-pipeline-caveats}, there may remain a 10\% systematic uncertainty in the calibration. 

\section{Results}
\label{results}

Figure~\ref{maps} shows the peak intensity maps of the lines in Table~\ref{lines}. We selected the H$_2$ S(1) line
to complete our study because it is located in channel 4 and presents similar excitation conditions and angular resolution as the sulfur lines relevant for this work. The high excitation rotational H$_2$ lines will be presented elsewhere (Sidhu et al., in prep). For reference, the positions HII, Atomic,  DF1, DF2, and DF3 as defined by \citet{Peeters2023} are indicated with crosses in all panels. The coordinates of these positions are listed in Table~\ref{fields}.The positions of the protoplanetary disks d203-504 and d203-506 \citep{Bally2000} are also marked with empty circles.
Different spatial distributions are observed according with the expected chemical and physical structure produced by the UV radiation impinging in the molecular cloud. The emission of the [S IV] 10.5 $\mu$m and [S III] 18.7 $\mu$m lines extends over all the mapped area with their intensities increasing by a factor of  $\sim$4 toward HII.  This spatial distribution is consistent with previous \textit{Spitzer} observations reported by \citet{Rubin2011} who showed that the emission of these lines extends to more than 10$'$ SW from $\Theta^1$ Ori C.  These authors interpreted this extended emission as coming from the fainter outer nebula usually called the Orion Veil \citep{Abel2004,Abel2006}. 
The emission of the H$_2$ S(1) line probes the warm molecular gas associated with the PDR formed between the interface between the HII region and the molecular cloud. This is a corrugated interface, which produces a series of bright H$_2$ S(1) rims (edge-on PDRs) emerging from an extended weaker emission (face-on PDRs)
\citep{Habart2023, Peeters2023}. The three brightest ridges are associated with the positions DF1, DF2, and DF3 (see Fig.~\ref{maps}). Weak H$_2$ S(1) emission coming from the face-on part of the PDR is detected in the region between these DFs, especially between DF2 and DF3.  The emission of the [S I] 25.249 $\mu$m line is weak and remains undetected toward HII, Atomic, and DF1. It is detected toward DF2 and DF3, and its emission remains high between them. In fact, its intensity smoothly increases beyond DF1 until DF3 in the mapped area (see Fig.~\ref{maps}). 

In Fig.~\ref{cuts} we show the normalized emission of the  [S IV] 10.5 $\mu$m, [S III] 18.7 $\mu$m,  [S I] 25.249 $\mu$m, and H$_2$ S(1) lines along the cut shown 
in Fig.~\ref{maps}. The emission of the [S IV] 10.5 $\mu$m and  [S III] 18.7 $\mu$m lines increases toward the HII region. 
The H$_2$ S(1) line is detected from the HII region to the molecular cloud, with peaks close to DF1, DF2, and DF3. A different distribution is observed in 
the [S I] 25.249 $\mu$m line whose emission smoothly increases toward the molecular cloud. Contrary to  H$_2$ S(1), the emission of  [S I] 25.249 $\mu$m remains high beyond DF3. 
This suggests that at least part of gas emitting in the [S I] 25.249 $\mu$m line is located deeper into the molecular cloud than that emitting in the H$_2$ S(1) line. 

\begin{figure}
\includegraphics[width=9cm]{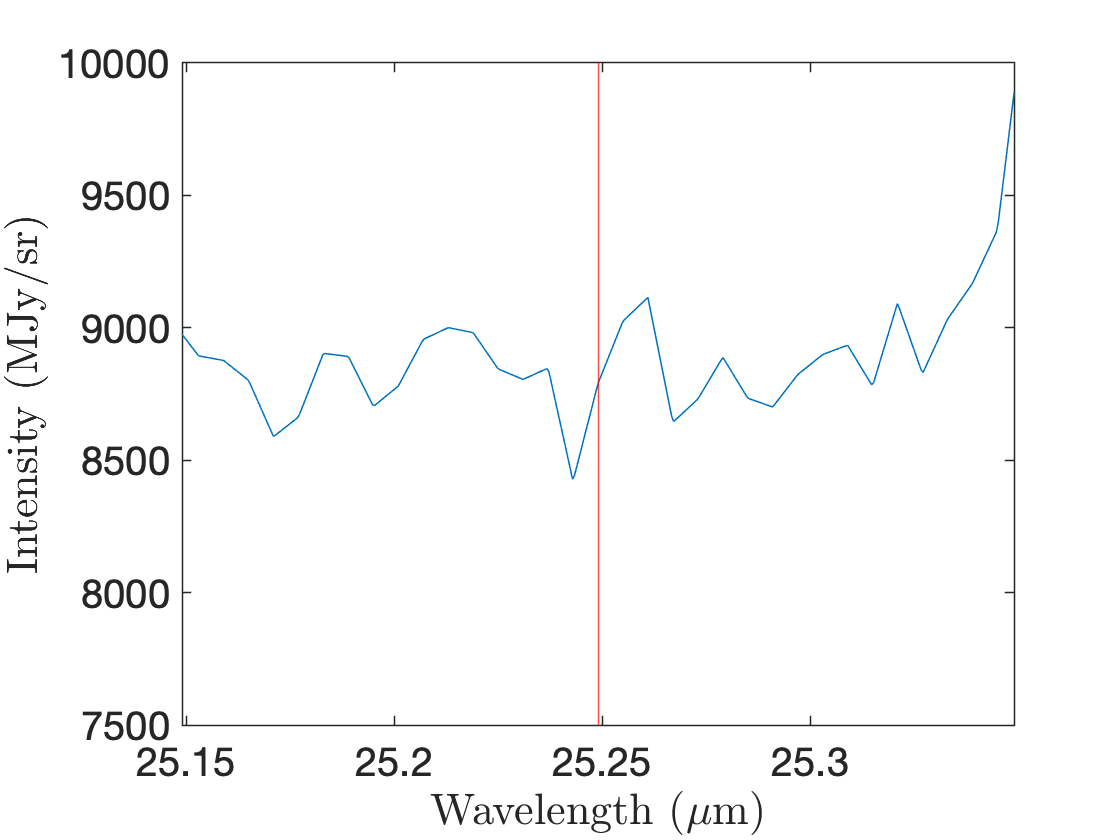}
\caption{ Spectrum of the d203-506 protoplanetary disk. The red line shows the wavelength of the [SI] line at 25.249 $\mu$m. 
A small feature at $\sim$ 25.257 $\mu$m is present but has an intensity lower than 2 $\times$ rms, precluding any firm detection.}
\label{proplyd}
\end{figure}

Two protoplanetary disks are present in our MIRI-MRS field of view: [BOM2000] d203-504, 
and [BOM2000] d203-506 \citep{Bally2000}.
These disks are well detected in JWST NIRCam  \citep{Habart2023, berne2024far}, NIRSpec \citep{Peeters2023, berne2024far}, 
and MIRI-MRS \citep{Berne2023, Zannese2024}.
There is no hint of detection of the [S I] 25.249 $\mu$m line toward d203-504. A weak peak at $\sim$ 25.249 $\mu$m  at the level of $\sim$ 2 $\times$ rms is found in the cut shown in Fig.~\ref{cuts} at the location of d203-506.  
In order to search more carefully for the [S I] 25.249 $\mu$m line in d203-506, we extracted the MIRI spectrum specifically for this source. We followed the method of \citet{Berne2023}, using an elliptical aperture fitted to the size of the source and the subtraction of a background nebular spectrum. 
The on-source spectrum is extracted from an ellipse centered on d203-506 at the position $\alpha=$ 5:35:20.357, $\delta=$ $-$5:25:05.81 with dimensions $l=0.52$'', $h=0.38$'' and a position angle PA=+33 degrees (trigonometric) with respect to north. The background nebular spectrum is taken at $\alpha=$ 5:35:20.370, $\delta=$ $-$5:25:04.97 using a circular aperture of radius $r=0.365$’’. The spectrum of d203-506 is obtained by subtracting the background spectrum from the on-source spectrum.
This spectrum is shown in Fig.~\ref{proplyd}. Although there is a slight excess of emission  around $\sim$25.257 $\mu$m in the spectrum toward d203-506, this excess is lower than 2 $\times$ rms and hence we cannot confirm a detection.  
Therefore, we can only provide an upper limit of $\leq$ 2.25 $\times$ 10$^{-5}$ erg s$^{-1}$ cm$^{-2}$ sr$^{-1}$ for the emission of the [S I] 25.249 $\mu$m line toward d203-506, which would imply  I ([S I] 25.249 $\mu$m) /I(H$_2$ S(1)) $\leq$0.13 in this silhouette disk using the data from \citet{Berne2023} for I(H$_2$ S(1)). A detailed analysis of the NIRSpec and MIRI spectroscopic data on this object will be the subject of a future paper. 
 
\section{Correction for extinction}
\label{fluxcorr}
\begin{figure}
\includegraphics[angle=0,scale=.34]{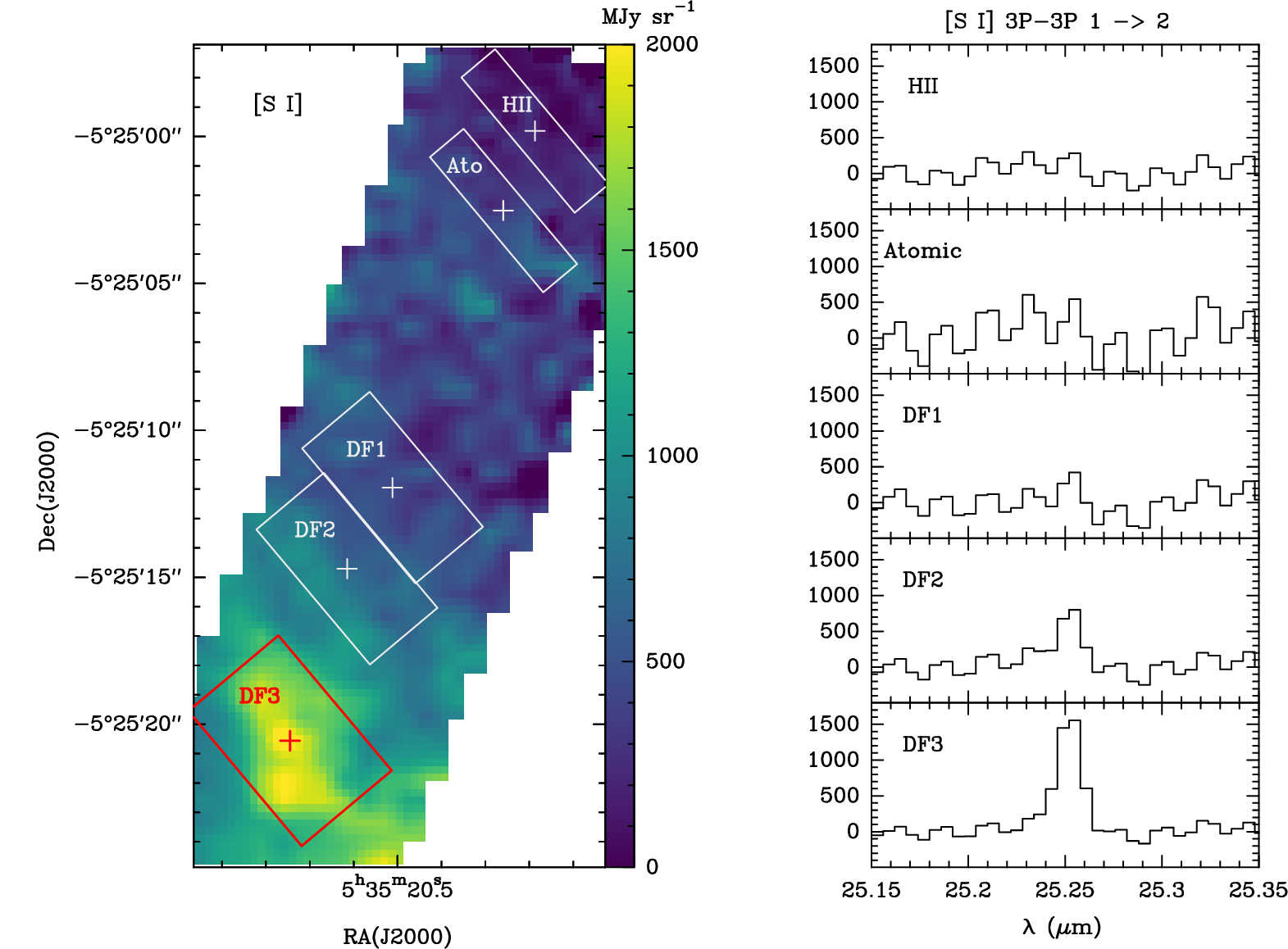}
\caption{Spectra of the [S I] 25.249 $\mu$m line in the Orion Bar. Left panel: Peak intensity map of the [S I] 25.249 $\mu$m line. We have drawn the rectangles used to extract the average intensity spectra. Right panel: Average intensity spectra of the [S I] 25.249 $\mu$m line in the rectangles shown in the left panel. }
\label{spectra}
\end{figure}

We extracted the average spectrum over five rectangular fields centered on the positions HII, Atomic, DF1, DF2, and DF3. The sizes and inclination angle of these rectangles are listed in Table~\ref{fields} and the obtained spectra are shown in Fig.~\ref{spectra} and Fig.~\ref{spectratodos}. The integrated line intensities obtained from these spectra are shown in Table~\ref{fluxes}. These intensities need to be corrected for the extinction between the gas emitting layer and the observer (A$^{los}_{\rm V}$) to obtain the emitted values. For that, we used the expression\begin{equation}
F_{corr} = F_{obs} \times \exp( \tau_\lambda), \\
\end{equation}
where $\tau_\lambda$ =  N$_{\rm H}$ $\cdot$ $\kappa_{abs}$,  with $\kappa_{abs}$ being the dust absorption coefficient per hydrogen nuclei. The value of N$_{\rm H}$ can be estimated from A$^{los}_{\rm V}$ using N$_{\rm H}/A^{los}_{\rm V}$=1.87 $\cdot$ 10$^{21}$ cm$^{-2}$ mag$^{-1}$ according to \citet{Joblin2018}.
The values of A$^{los}_{\rm V}$  and $\kappa_{abs}$ suffer from large uncertainties. 
The extinction between the emitting gas and the observer depends on the frequency and the projected distance from $\Theta^1$ Ori C (see the scheme in  Fig.~\ref{fig1}). 
\citet{Peeters2023} derived the extinction in the ionized gas using the H I recombination lines. They referred to this value 
as A$_V$(foreground) (see Table~\ref{fields}) and we used it  as A$^{los}_{\rm V}$ for the extinction correction of the ions emission.
The visual extinction produced by the molecular bar was estimated by the same authors using the H$_2$ ro-vibrational near-infrared lines and two different assumptions: 
(i) the dust responsible for the visual extinction is in front of the H$_2$ emitting layer (A$_{\rm V}$ (bar)$^1$ in Table~\ref{fields}); and (ii) the intermingled formalism, which assumes that the dust is mixed with the gas emitting in the H$_2$ lines (see A$_{\rm V}$ (bar)$^2$ in Table~\ref{fields}). Since these assumptions 
correspond to two limiting cases,  A$_{\rm V}$ (bar)$^1$ and   
A$_{\rm V}$ (bar)$^2$ can be considered as lower and upper limits to the visual extinction produced by the molecular bar. 
As discussed in Sect.~\ref{results}, the emission 
of the  [S I]  25.249~$\mu$m  line comes from the molecular gas.
Accordingly, we estimated lower and upper limits to the H$_2$ S(1) and  [S I]  25.249~$\mu$m extinction-corrected intensities using A$^{los}_{\rm V}$ = A$_{\rm V}$ (foreground) $+$A$_{\rm V}$ (bar)$^1$ and A$^{los}_{\rm V}$ = A$_{\rm V}$ (foreground)$+$A$_{\rm V}$ (bar)$^2$ (see Table~\ref{corrs}).
To calculate $\kappa_{abs}$, we used the dust opacities tabulated by \citet{Ossenkopf1994} for bare grains, Mathis distribution, and assumed dust-to-gas ratio of 0.01. The line intensities thus corrected are shown in Table~\ref{corrs} and are the ones used to compare with our models. The correction for extinction is less than 33\% for the H$_2$ S (1) and [S I] 25.249 $\mu$m lines in all fields. Therefore, we do not expect that the uncertainties in A$^{los}_V$ and the optical dust properties have a significant impact in our conclusions.

 \begin{table*}
\caption{{Abundances of sulfur ions}.}
\label{SulfurIonsv2}
\begin{tabular}{lcc|ccc|c|ccc|c}\\
\hline
\noalign{\smallskip}
ID &  T$_e$  &    n$_e$  & E$_{\rm em}$ ([SII])$^a$     &   E$_{\rm em}$ ([S III])$^b$   &   E$_{\rm em}$ ([S IV])$^c$   &   E$_{\rm em}$  (Pf$\beta$)$^d$  
& S$^+$/H$^+$ & S$^{++}$/H$^+$ & S$^{3+}$/H$^+$ &  [S/H]$_{ion}$   \\
    &    (K)     & (cm$^{-3}$)   &  \multicolumn{3}{|c|}{(erg s$^{-1}$)}  &  \multicolumn{1}{|c|}{(erg cm$^{-3}$ s$^{-1}$)} &   \multicolumn{3}{|c|}{} &  \\
     \hline
H II      &  9000     &   3330    & 
4.85(-17) &
3.64(-17)  &  
1.70(-16)  &  
2.42(-27)   &  
1.30(-6) &
7.02(-6)   &  
8.09(-8)  &  
8.40(-6)    \\
Atomic       &  8690     &   3190  &
4.40(-17)& 
3.58(-17)    &   
1.65(-16)   &   
2.34(-27)     &  
1.09(-6)&
7.08(-6)   &  
1.63(-7)  &
8.33(-6)   \\
DF 1     &  8130    &   2050 &
2.76(-17)  &   
2.73(-17)    &    
1.16(-16)   &   
2.54(-27)  &
1.04(-6)   &
6.80(-6)   &   
2.54(-7)  & 
8.09(-6)  \\
DF 2    &  8160    &   1970 &
2.70(-17) &   
2.66(-17)    &
1.12(-16)   &
2.54(-27)   &  
1.08(-6)   &
6.84(-6)   &   
2.65(-7)  &   
8.19(-6)  \\
\hline  
\end{tabular}

\noindent
$^a$ Emissivity of the [SII] 6731 \AA\ line per S$^{+}$ atom; 
$^b$ Emissivity of the [S III] 18\,$\mu$m line per S$^{++}$ atom;  
$^c$ Emissivity of the [S IV] 10\,$\mu$m line per S$^{3+}$ atom;  
$^d$ Emissivity of the Pf$\beta$ line per n$_{\rm H^+}$$\times$n$_e$\\
\noindent
Notation: 4.85(-17) = 4.85$\times$10$^{-17}$

\end{table*}

\section{Sulfur abundance in the ionized gas}
\label{ion}

As noted in Sect.~\ref{results}, the emissions of the [S III] 18.71\,$\mu$m and [S IV] 10.51\,$\mu$m lines  extend over several arcminutes further from the IF. 
Both the extension and the emission profiles of the sulfur ionized lines can be explained if a significant fraction of their emissions come from the fainter foreground ionized gas.
This conclusion is confirmed by the \textit{Spitzer} data  reported by \citet{Rubin2011}, who detected
emission of the [S III] 18.71\,$\mu$m and [S IV] 10.51\,$\mu$m lines at distances of more than 10$'$ from $\Theta^1$ Ori C. In the following, we estimate the S/H ratio in HII, Atomic, DF1, and DF2 fields based on JWST data and previous optical and near-infrared observations.

The S/H ratio in the ionized gas is calculated by adding up the amount of S$^{3+}$, S$^{++}$ and S$^+$ along the line of sight
and comparing it with the amount of atomic hydrogen as derived from the Pfund 7$\rightarrow$5 recombination line at 4.654\,$\mu$m (also called Pf$\beta$) taken from 
\citet{Peeters2023}. 
In order to estimate each species (H$^+$, S$^{++}$, S$^{3+}$) column density, we needed to calculate the line emissivities per particle along the line of sight (hereafter, E$_{\rm em}$). The emissivity of each line can be calculated as  E$_{\rm em}$ = $f_u \times  E_u \times  A_{ij}$, where $f_u$ is the fraction of ion density in the upper level, and $E_u$ and  $A_{ij}$ are the upper level energy and the Einstein spontaneous emission coefficient, respectively.
We obtained the physical conditions in our reference fields from the images of T$_e$ and n$_e$ derived from 
[S III] 6312\,\AA/9069\,\AA\ and [SII] 6731\,\AA/6716\, \AA\ line ratios by \citet{Weilbacher2015} (Figs. 5 and 6 of this paper). 
\citet{Storey1995} tabulated the Pfund 7$\rightarrow$5 recombination line emissivity  as a function of  T$_e$ and (n$_{\rm H^+}$ $\times$ n$_e$).
For each field, we adopted the value of  E$_{\rm em}$ (Pf$\beta$) corresponding to its physical conditions.
We calculated the values of $N_u$  for the [S III] 18.71\,$\mu$m and [S IV] 10.51\,$\mu$m 
considering only collisions with electrons. The values of  E$_{\rm em}$ thus calculated are shown in
Table~\ref{SulfurIonsv2}.
Then, using the calculated emissivities and the extinction-corrected line intensities (Table~\ref{corrs}), we estimated N(S$^{++}$), N(S$^{3+}$), and N(H$^+$), and their ratios with respect to N(H$^+$) in HII, Atomic, DF1, and DF2
using
\begin{gather}
N(\text{S}^{+}) = I_{\rm obs} (\text{[SII]}) \times 4 \pi /  E_{\rm em}(\text{[SII]}) \\ 
N(\text{S}^{2+}) = I_{\rm obs} (\text{[S III]}) \times 4 \pi /  E_{\rm em}(\text{[S III]}) \\
N(\text{S}^{3+}) = I_{\rm obs} (\text{[S IV]}) \times 4 \pi /  E_{\rm em}(\text{[S IV]}) \\
N(\text{H}^+) = I_{\rm obs} (\text{Pf} \beta) \times 4 \pi /  [E_{\rm em}(\text{Pf} \beta ) \times n_e].
\end{gather}

We needed to know N(S$^+$) in order to account for the total sulfur budget in ionized gas phase. Since there are no near-infrared [S II] lines arising from the ground state, we used 
the [SII] 6731\,\AA\ map from \cite{Weilbacher2015} along with the emissivity 
calculated with CHIANTI (v10.1.2; \citealp{DelZanna2021}) to obtain N(S$^{+}$). The N(S$^+$)/N(S$^{++}$) ratios obtained with our calculations
range from 0.15 to 0.18, in agreement with \citet{Rubin2011}, who find 0.19 as an upper limit. 
The elemental abundance of sulfur in the ionized gas is then calculated 
as [S/H]$_{ionized}$ = [N(S$^{3+}$)+ N(S$^{++}$)+N(S$^+$)]/N(H$^+$).
With these assumptions the S/H ratios are estimated to be 
8.40, 8.33, 8.09, and 8.19$\times 10^{-6}$, respectively, in the HII, Atomic, DF1, and DF2 regions (see Table~\ref{SulfurIonsv2}).
These values are roughly consistent with the results of \citet{Rubin2011} 
who determined S/H = (7.68$\pm$0.25)$\times$10$^{-6}$ in the Orion Veil based on near-infrared lines.

The estimated value of S/H toward HII is a factor of $\sim$1.3-1.8 lower than that derived for the M~42 nebula by previous authors.  
\citet{Esteban2004} observed a region near the hot star $\Theta^1$ Ori, and measured S/H $\sim$ 1.65 $\times$ 10$^{-5}$ when temperature 
fluctuations are included and $\sim$ 1.15 $\times$ 10$^{-5}$ when they are not,  while \citet{McLeod2016} reported a value of 1.1 $\times$ 10$^{-5}$ toward the Orion Bar. 
\citet{Daflon2009} estimated the sulfur abundance based on the photospheric lines of  a sample of ten B main-sequence stars of the Orion association and
obtained S/H = (1.41$\pm$0.17) $\times$ 10$^{-5}$, which is consistent with the solar value and that of meteorites \citep{Asplund2006}. The estimated S/H ratio in the 
Orion Veil is, therefore, a factor of $\sim$2 lower than the solar value and that obtained in the M~42 nebula, but still within the uncertainties in this kind of calculation.
This discrepancy could originate from the simplicity of our model, which assumes uniform physical
conditions along the line of sight, thus neglecting the structure of the ionized gas.

\begin{figure}
\includegraphics[angle=0,scale=.7]{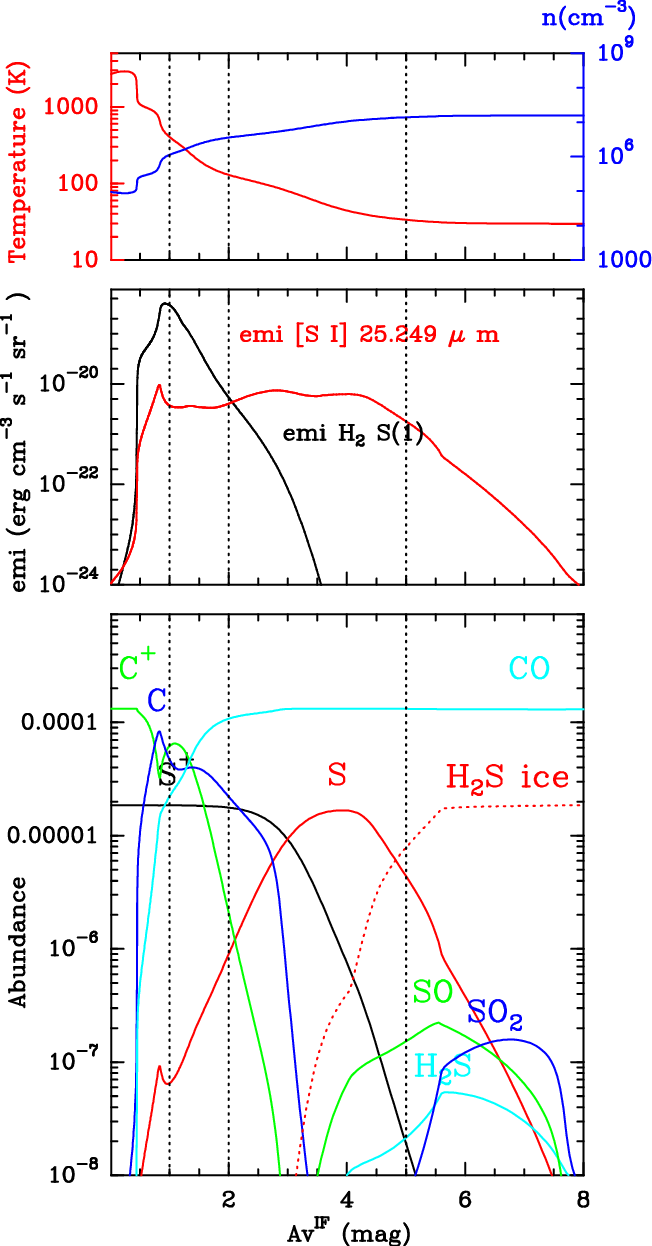}
\caption{Predictions of the Meudon PDR code for the reference model. From top to bottom: Physical conditions, volume emissivities of the H$_2$ S(1) and [S I] 25.249 $\mu$m lines, and chemical abundances relative to hydrogen nuclei as a function of the visual extinction from the HII region (A$^{IF}_V$). The volume emissivity is defined as $emi $ = $n \times f_u \times  E_u \times  A_{ij} /(4 \times \pi)$, where $n$ is the gas density, $f_u$ is the fraction of particles in the upper level, and $E_u$ and  $A_{ij}$ are the upper level energy and the Einstein spontaneous emission coefficient. The transition of S$^+$/S takes place at a  visual extinction, A$^{IF}_V$ $\sim$ 3.0 mag, where the gas temperature drops below 100 K and densities are  $\sim$10$^7$ cm$^{-3}$.  The local volume emissivity of the  [S I] 25.249 $\mu$m line remains high deeper into the molecular cloud, until A$^{IF}_V$$\sim$ 5 mag. We indicate A$^{IF}_V$ = 1, 2, and 5 mag with dashed vertical lines. 
}
\label{chemi1}
\end{figure}

\section{Chemical modeling: Atomic sulfur (S) }
\label{model}

In this section we model the emission of the H$_2$ S(1) and [S I] lines in order to determine sulfur depletion in the Orion Bar using the
 Meudon PDR code\footnote{https://pdr.obspm.fr}.
This code can simulate very detailed micro-physical processes for a given value of the incident UV field and a 1D density structure. 
As output, it provides  the gas and dust temperatures as well as the chemical abundances at each position in the 1D slab, and also performs line and continuum 
radiative transfer to calculate the integrated emerging fluxes for different inclination angles along the line of sight  
\citep{Goicoechea2007, Gonzalez2008, Bourlot2012, Bron2014, Bron2016}.
Atomic and molecular properties can also be viewed as important parameters of the code. We carefully updated the chemical network, focusing specifically 
on sulfur species, following the recent studies of \citet{Bulut2021} and \citet{Fuente2019,Fuente2017-B1} for the CS + O, SO + OH and O$_2$ + S reactions. In addition, we introduced the experimental photodissociation sections of SH, CS and H$_2$CS recently implemented by \citet{Hrodmarsson2023} as well as the new dissociative recombination rate 
of SH$^+$ by \citet{Boffelli2023}. The final network consists of 3052 gas reactions including photo-reactions. In addition, we added new photo-destruction cross-sections to the Meudon database for SO, SO$_2$, CS, O$_2$, HCN, and HNC, in place of approximate analytical fits, to obtain more accurate predictions. These cross sections were retrieved from the Leiden database at: https://home.strw.leidenuniv.nl/~ewine/photo/index.html, as described in \citet{Hrodmarsson2023}  and \citet{Heays2017}. References to the original sources are given in these papers and in the respective data files. These data were then resampled to fit the needs of the PDR code while keeping the optimal resolution.

Pure gas-phase chemistry is insufficient to explain the abundances of gaseous hydrogen sulfide in the Orion Bar \citep{Goicoechea2021}. A plausible formation channel is provided by surface reactions between accreted atomic sulfur and hydrogen. We included in our chemical network the basic sulfur
and oxygen surface chemistry described in \citet{Goicoechea2021} to account for sulfur accretion and hydrogenation on grain surfaces.
In particular, we considered the adsorption and desorption of H, H$_2$, O,  O$_2$, OH, H$_2$O, CO, S, SH, and H$_2$S and the formation of H$_2$, H$_2$O, and H$_2$S on
the grain surfaces.  
 
Inelastic collisions with electrons, helium, atomic hydrogen and ortho- and para-H$_2$ need to be taken into account to calculate the excitation of neutral atomic sulfur. 
We adopted the collision rates with electrons calculated by \citet{Tayal2004}. For atomic hydrogen, we used those calculated by \citet{Yan2023}. For helium, we used the collision rates computed by \citet{Lique2018}. Motivated by the present study, inelastic collision rates with ortho- and para-H$_2$ were calculated as explained in the following section.

\begin{table*}
\caption{Parameters of our reference model.}
\label{refmod}
\begin{tabular}{lll}\\
\hline \hline
Parameter  &   Value      &    Reference   \\
\hline 
G$_0$ at IF                                     &   3.1 $\times$ 10$^4$  Mathis field   & \citet{Joblin2018}   \\          
G$_0$ at the back side                   &  3.1 $\times$ 10$^3$  Mathis field   & \citet{Joblin2018}  \\                                  
A$_V$                                             &   20 mag                                           & \citet{Joblin2018} \\                  
Inclination angle ($i$)                      &   60$^\circ$                                       & \citet{Joblin2018}  \\                                                                                        
Thermal pressure (P$_{\rm th}$)    &    2.8 $\times$ 10$^8$  K cm$^{-3}$  & \citet{Joblin2018}   \\        
R$_V$=A$_V$/E(B-V)                    &   5.6                                                  & \citet{Marconi1998, Witt2006}   \\                                   
N$_H$/E(B-V)                                 &   1.05 $\times$ 10$^{22}$ cm$^{-2}$ mag$^{-1}$   &  \citet{Joblin2018}  \\
$\zeta_{H_2}$                                 &   5 $\times$ 10$^{-17}$ s$^{-1}$                            &  \citet{Joblin2018}  \\
Mass grain/mass gas                     &  0.01                                                                    &   \\
Grain size distribution                     & $\propto a^{-3.5}$                                                 & \citet{Mathis1977} \\
Min. grain size                                &  3 $\times$ 10$^{-7}$ cm    &  \citet{Joblin2018}  \\
Max. grain size                               &  3 $\times$ 10$^{-5}$ cm    &  \citet{Joblin2018}  \\
C/H                                                 &   1.32 $\times$ 10$^{-4}$   & \citet{Savage1996} \\
O/H                                                 &   3.19 $\times$ 10$^{-4}$   & \citet{Meyer1998} \\
S/H                                                 &   1.86 $\times$ 10$^{-5}$    & \citet{Savage1996} \\
                                                       &   1.40 $\times$ 10$^{-5}$    & \citet{Asplund2006, Daflon2009, GoicoecheaCuadrado2021} \\
N/H                                                 &   7.50 $\times$ 10$^{-5}$    & \citet{Meyer1997}  \\
\hline \hline
\end{tabular}
\end{table*}

\subsection{Sulfur -- H$_2$ collision rates}

The rate coefficients 
for fine structure excitation of neutral atomic sulfur in collisions with ortho-
and para-H${}_2$ 
for temperatures up to 2000~K were obtained
from the corresponding cross sections assuming a Maxwellian
velocity distribution.
The cross sections utilized were calculated using a quantum-mechanical
scattering theory detailed previously by \citet{Yan2023} with
some modifications for the treatment of molecular hydrogen.
Potential energy surfaces for the 
electronic states arising from the interaction of S$(^3P)$ and H${}_2$ were
calculated using computational quantum chemical methods
with reliable basis sets. 
The methodology was validated by carrying out additional
analogous calculations for the O-H${}_2$ system
and good agreement was obtained
with previous studies~\citep{Flower1990,Jaquet1992}.
A detailed description of the S-H$_2$ collisional calculations will be presented elsewhere (P.G. Yan and J.F. Babb, in prep.).

\subsection{Chemical abundances: Reference model}

DF3 is the position where the emission of the  [S I] 25.249 $\mu$m line is highest. This dissociation front is also the most intense in the far-infrared CO
lines and in the ground rotational lines of H$_2$ as shown by \citet{Joblin2018}.  These authors used the Meudon PDR code to explain the set of observations provided by the \textit{Herschel} Space Observatory in this PDR. The best fit was obtained assuming a plane-parallel 1D isobaric slab with incident UV field, G$_0$~=~3.1 $\times$ 10$^4$ Mathis field, thermal pressure, P$_{th}$~=~2.8 $\times$ 10$^8$ cm$^{-3}$ K, and cosmic ray ionization rate per H$_2$ molecule, $\zeta_{\rm H_2}$~=~5 $\times$ 10$^{-17}$ s$^{-1}$. The thickness of the slab was assumed to be A$_{\rm V}$=20 mag and the illumination on the back side ten times lower than that in the side facing the HII region. An inclination angle, $i$=60$^{\circ}$ provided the best fit with the observations. 
The adopted elemental abundances were: C/H=1.32 $\times$10$^{-4}$, O/H=3.19 $\times$10$^{-4}$, N/H=7.50 $\times$10$^{-5}$, and S/H=1.86 $\times$10$^{-5}$. The value of sulfur abundance is the most uncertain in this list. 
Recent calculations of the sulfur abundance in the Orion nebula are values closer to S/H=1.40 $\times$10$^{-5}$ \citep{Daflon2009,GoicoecheaCuadrado2021}, which is consistent with the value in Solar System \citep{Asplund2006}. Taking into account the uncertainties involved in the estimates of the sulfur abundance, we kept S/H=1.86 $\times$10$^{-5}$ in our modeling, which is consistent with the upper limit derived by \citet{GoicoecheaCuadrado2021} in this PDR and provides the best fit to the observations. Hereafter, this set of input parameters is referred to as $"$reference model$"$ (see Table~\ref{refmod}) and will be used to fit the new JWST observations.  

Figure~\ref{chemi1} shows the gas physical conditions and the fractional abundances of the most abundant carbon and sulfur species as a function of the visual extinction from the IF as predicted by the reference model. We recall that the visual extinction from the IF (A$^{IF}_V$) is different from that along the line of sight (A$^{los}_V$) used in Sect.~\ref{fluxcorr} (see the scheme in Fig.~\ref{fig1}). The C$^+$ and neutral C carry most of the carbon in gas phase until a visual extinction, A$^{IF}_V$ $\sim$ 1.3 mag, where CO becomes the most abundant species. The gas temperature is $\sim$ 200 K and densities are approximately a few times 10$^6$ cm$^{-3}$ at this visual extinction. Most of the sulfur is in the form of S$^+$ and the fractional abundance of S is $\sim$10$^{-7}$. The transition of S$^+$/S takes place at a visual extinction, A$^{IF}_V$ $\sim$ 3 mag, where the gas temperature drops below 100 K and densities are increasing to $\sim$10$^7$ cm$^{-3}$.  Neutral atomic sulfur remains the main sulfur reservoir until A$^{IF}_V$ $\sim$5 mag. Beyond  $\sim$5 mag, most of the sulfur is locked in solid H$_2$S and the most abundant sulfur species in gas phase are SO, SO$_2$, and H$_2$S. 

The emission of a given line depends on the abundance of the carrier and local physical conditions. 
The contribution of each parcel of gas to the observed emission is given by the volume emissivity calculated as  $emi $ = $n \times f_u \times  E_u \times  A_{ij} /(4 \times \pi)$, where $n$ is the gas density and $f_u$ is the fraction of a given species in the upper transition level, and $E_u$ and  $A_{ij}$ are the upper level energy and the Einstein spontaneous emission coefficient.
In Fig.~\ref{chemi1}, we show the local emissivities of the H$_2$ S(1) and [S I] 25.249~$\mu$m lines as a function of A$^{IF}_V$. 
While the emissivity of the H$_2$ S(1) line has a narrow peak at A$^{IF}_V$ $\sim$ 1 mag and drops by several orders of magnitude at A$^{IF}_V$$>$ 2 mag, the emissivity of the  [S I] 25.249 $\mu$m line has a shallow peak at the same position but remains almost constant, and higher than 10$^{-21}$ erg cm$^{-3}$ s $^{-1}$ sr$^{-1}$,  until A$^{IF}_V$$\sim$ 5 mag. 
This prediction is consistent with the morphology observed in the Orion Bar, where the H$_2$ S(1) rotational line is mainly tracing the edge-on surface of the PDR, while the  [S I] 25.249 $\mu$m line arises in the molecular cloud. 

We carried out line transfer calculations to predict line intensities and compare with observations on quantitative grounds. The excitation of the  [S I]  25.249 $\mu$m line is dominated by collisions with ortho- and para-H$_2$ for A$^{IF}_V$$>$ 1 mag where the gas is mainly in molecular form. 

\begin{table*}
\caption{Comparison between predicted (for i=60$^\circ$) and observed intensities (erg cm$^{-2}$ s$^{-1}$ sr$^{-1}$).$^*$}
\label{compa}
\begin{tabular}{cccl}\\
\hline 
\noalign{\smallskip}
 I[H$_2$ S(1)]  &   I( [S I]~25.249~$\mu$m)  &  $\frac{{\rm I}({\rm [S I}]~25.249~\mu m)}{{\rm I}[{\rm H_2}~{\rm S(1)}]}$   &  Models$^1$  \\ \hline 
1.16(-3)           &   3.72(-6)   &   0.003    &    Model 1 (P=2.8 $\times$ 10$^7$ K cm$^{-3}$)  \\ 
7.61(-4)           &   2.80(-5)   &   0.04      &    Model 2 (P=2.8 $\times$ 10$^8$ K cm$^{-3}$)  \\ 
4.86(-4)           &   4.58(-4)   &   0.94      &    Model 3 (P=2.8 $\times$ 10$^9$ K cm$^{-3}$)  \\ 
1.98(-4)           &   1.37(-5)   &   0.07      &    Model 4 (G$_0$=3.1 $\times$ 10$^3$ Mathis field) \\   
1.12(-3)           &   2.44(-4)   &   0.21      &    Model 5 (G$_0$=3.1 $\times$ 10$^5$ Mathis field) \\   
7.57(-4)           &   2.69(-5)   &   0.03      &    Model 6 ($\zeta_{H_2}$=5 $\times$ 10$^{-16}$ s$^{-1}$)  \\ 
1.11(-3)           &   4.65(-5)   &   0.04      &    Model  8 (N$_{\rm H}$/E(B-V)=1.6 $\times$ 10$^{22}$ cm$^{-2}$  mag$^{-1}$) \\ 
7.23(-4)           &   2.00(-7)   &   0.0004  &    Reference  (S/H=1.86 $\times$ 10$^{-7}$) \\ 
8.95(-4)          &    4.23(-5)   &   0.05   &       Reference   (S/H=1.86 $\times$ 10$^{-5}$) \\  
\hline
8.62(-4)          &  1.19(-4)    &   0.14    &   DF3    \\
\hline 
\end{tabular}

\noindent
Notation: 8.62(-4) = 8.62 $\times$ 10$^{-4}$ \\
\noindent
$^*$ In this table, only the model parameters relevant for the discussion in Sect.~\ref{pred} are shown. The complete set of input parameters are given in Table~\ref{refmod} and Table~\ref{mod}.
\end{table*}

\subsection{Predicted line intensities}
\label{pred}

 \subsubsection{Reference model}
Reference model provides a good guess for the physical parameters in DF3. As commented above, it is the best fitting to the far-infrared CO lines and in the ground rotational lines of H$_2$ as observed with \textit{Herschel} and reported by \citet{Joblin2018}. 
In Table~\ref{compa}, we compare the intensities of the H$_2$ S(1) and [S I]  25.249~$\mu$m lines predicted by reference model with observations. For the computation of these lines we assumed an inclination angle, $i$ = 60$^{\circ}$. Reference model predicts the intensities of the H$_2$ S(1) and   [S I]  25.249~$\mu$m lines within a factor of $\sim$3. Moreover, the predicted I( [S I]   25.249~$\mu$m ) / I( H$_2$ S(1) ) is a factor of $\sim$3 lower than the observed one, which is a reasonable agreement with observations taking into account the uncertainties in our modeling.  It should be noticed that a higher value of $i$ would produce higher line intensities but would not change the I( [S I]   25.249~$\mu$m ) / I( H$_2$ S(1) ) ratio. We recall here that our model assumes a S/H = 1.86 $\times$10$^{-5}$, which is consistent with the upper limit derived by \citet{GoicoecheaCuadrado2021} on the basis of the sulfur recombination lines. In some sense, this is an upper limit to the possible sulfur elemental abundance. 
This elemental abundance is  in agreement within a factor of 2 with previous results based on ionized and recombination lines \citep{Daflon2009,Rubin2011,GoicoecheaCuadrado2021}. We do not consider that a factor of 2 is significant within the uncertainties of this kind of modeling.  Our results support the interpretation that sulfur remains undepleted along the ionized, atomic, and molecular phase in the Orion Bar.

\subsubsection{Sensitivity to input parameters}
Next we explored the sensitivity of the predicted [S I]  25.249~$\mu$m and H$_2$ S(1) intensities to different physical parameters. For this purpose, we ran the grid of 1D models described in Table~\ref{mod}. The computed values of I (  [S I]  25.249~$\mu$m ), I( H$_2$ S(1) ), and I (  [S I]  25.249~$\mu$m )/I( H$_2$ S(1) ) ratios are listed in Table~\ref{compa}.
In models 1 to 5, we investigated the impact of the thermal pressure and the incident UV field on the computed intensities. 

The value of I ([S I]  25.249~$\mu$m) increases almost linearly with P$_{\rm th}$ (see models 1 to 3 in Table~\ref{compa}).
This is not unexpected because of the high critical density (n$_{\rm H}$ $\sim$ 10$^5$ cm$^{-3}$) and upper level energy of this transition (see Table~\ref{lines}). 
However, I (H$_2$ S(1)) decreases with increasing P$_{\rm th}$ because higher excited ro-vibrational levels of H$_2$ become populated.  This means that higher values of 
P$_{\rm th}$ would produce higher values of I ( [S I]  25.249~$\mu$m)  and  I( [S I]   25.24~$\mu$m ) / I( H$_2$ S(1) ). 

Increasing G$_0$ would lead to higher values of both,  I( [S I]  25.249~$\mu$m ) and I( H$_2$ S(1) ) (see models 2, 4 and 5 in Table~\ref{compa}). One important result is that  I ( [S I]  25.249~$\mu$m) $<$10$^{-5}$ erg cm$^{-2}$ s$^{-1}$ sr$^{-1}$for G$_0$ $<$ 10$^4$ Mathis field. This limits the detectability of the [S I]   25.249~$\mu$m line to the PDRs associated with the HII regions formed by massive stars. 

To interpret our observations, it is interesting to explore the influence of other parameters such as $\zeta_{\rm H_2}$, N$_{\rm H}$/E(B-V) ratio, and S/H. 
The comparison of models 2 and 6 shows that increasing the value of $\zeta_{\rm H_2}$ by a factor of 10
produces minor changes in the intensities of both lines. Therefore, we do not expect that possible uncertainties in the value of  $\zeta_{\rm H_2}$ would affect our conclusions. 

The dust extinction curve is not well known, and several values of N$_{\rm H}$/E(B-V) have been used in the literature \citep{Cardelli1989, Joblin2018, Habart2023}. In particular, \citet{Habart2023} used  N$_{\rm H}$/E(B-V) = 1.6 $\times$ 10$^{22}$ cm$^{-2}$ mag$^{-1}$ from \citet{Cardelli1989}. In order to test the impact of this parameter we ran model 7. Changing N$_{\rm H}$/E(B-V) ratio  (models 2 and 7) have a significant impact on I ( [S I]   25.249~$\mu$m ) and I ( H$_2$ S(1) )  but the I( [S I]   25.249~$\mu$m ) / I( H$_2$ S(1) ) ratio is little affected. 

As expected, varying S/H has a large impact of the intensity of the  [S I]   25.249~$\mu$m line, which scales almost linearly with S/H, but has no significant impact of the intensity of the H$_2$ S(1) line (models 2 and 8). Therefore, the  I(  [S I]   25.249~$\mu$m ) / I( H$_2$ S(1) ) ratio increases with S/H and can be used as a tracer of S/H in PDRs as long as P$_{\rm th}$ and G$_0$ are known. Therefore, the study of neutral atomic sulfur in PDRs through the observation of the [S I] 25.249 $\mu$m fine-structure line can provide a good estimate of sulfur depletion in this kind of region.

\subsection{Limitations of our modeling}
In this section, we revise the uncertainties in the sulfur chemistry that could affect our results. The uncertainties in the adsorption and desorption processes of sulfur species, in particular atomic sulfur, on the grain surfaces could have a significant impact on the predicted chemical abundances. Indeed, the binding energies of the sulfur species on the grain surfaces are not fully understood yet. \citet{Perrero2022} calculated the binding energies of S, SH, and H$_2$S in amorphous and crystalline ice. They found that each species experience different binding energies depending on its position in the ice structure.  In general,  binding energies are higher in crystalline than in amorphous water ice. In amorphous ice, they found a range of values for the binding energy of a given species with variations of up to a factor of 2 depending on the adsorption site. Binding energies  previously reported by  \citet{Wakelam2017} and \citet{Das2018} are placed within this range. \citet{Perrero2024} estimated the binding energies of sulfur species on bare silicate grains. Their results show that binding energies in bare grains are a factor of 2$-$4 higher than those in amorphous ice. In our model, we adopted BE (H$_2$S) =  BE (SH) = 2700 K, which are at the lower end of the range values published by \citet{Perrero2022} in amorphous ice, and the ones reported by \citet{Wakelam2017}. For atomic sulfur, we adopted BE(S)=1800 K, close to the minimum value reported by \citet{Perrero2022} in amorphous ice and consistent with the experiments by \citet{JimenezEscobar2011}. These values were also used by \citet{Cazaux2022} to investigate the formation of allotropes on grain surfaces.  Therefore, the adopted binding energies lie at the lower end of the possible values, hence maximizing the amount of sulfur in gas phase.
 
Nonthermal processes can also contribute to release sulfur species to the gas phase in PDRs.
Unfortunately, only the H$_2$S photo-desorption yield in pure H$_2$S ice has been estimated thus far.  Based on laboratory experiments, \citet{Fuente2017-S2H} determined the photo-desorption yield of H$_2$S to be 1.2 $\times$ 10$^{-3}$ per molecule and incident photon. For the other sulfur species, we assumed 1 $\times$ 10$^{-4}$ per particle and photon following \citet{HasegawaHerbst1993}. Laboratory experiments to measure the photo-desorption yields for all sulfur species would be desirable in order to have a more accurate description of the gas-dust interaction.

To test the influence of the adopted binding energies and photo-desorption yields on the predicted intensity of the [S I] 25.249 $\mu$m line, we ran a pure gas-phase model (i.e., neglecting adsorption and desorption on grain surfaces) using the input parameters in Table~\ref{refmod}. As expected, the I( [S I] 25.249 $\mu$m ) and I( [S I]   25.249~$\mu$m ) / I( H$_2$ S(1) ) ratios increase but  the model falls short, by a factor of $\sim$2, in the prediction of the observed intensity. 

The discrepancy between model and observations can also be due to the assumed physical structure. As commented above, the intensity of the [S I] 25.249 $\mu$m line is very sensitive to assumed thermal pressure. High angular resolution images ($\sim$ 1\arcsec) of the CO J=3$\rightarrow$2 line revealed the presence of fragmentation and  photo-evaporation flows in the Orion Bar \citep{Goicoechea2016}.  \citet{Bron2018} developed a 1D hydrodynamical PDR code coupling hydrodynamics, extreme-ultraviolet and FUV radiative transfer, and time-dependent thermochemical evolution to simulate an UV-illuminated molecular cloud that is evaporating into a surrounding low-pressure medium. 
They found that, although moderate pressure gradients can develop in this scenario, isobaric PDR models are a better approximation to the structure of photoevaporating PDRs than constant-density PDR models. The reference model is therefore our best approximation to the physics and chemistry in the Orion Bar. Though small departures in the physical structure from the isobaric case, for example variations in the density, are not unreasonable and could help reconcile model predictions and observations.

\section{Discussion}

We used high sensitivity JWST observations  to estimate the sulfur abundance in the ionized and warm molecular layer of the Orion Bar. The detection of the [S I] 25.249 $\mu$m line shows that sulfur is undepleted in the warm molecular gas. 

The lack of sulfur depletion in the warm molecular gas of the Orion Bar challenges our understanding of sulfur chemistry. In order to explain the high value of sulfur depletion estimated in dark clouds (see, e.g., \citealp{Fuente2019}), it has been proposed that most of the sulfur can be locked in (semi-)refractories such as iron sulfide and sulfur allotropes in these cold clouds \citep{JimenezEscobar2011, JimenezEscobar2012, Fuente2019, Kama2019, Shingledecker2020, Cazaux2022}. Refractory material presents sublimation temperatures higher than a few hundred K and can only be evaporated in extremely hot regions (T$\sim$ 1000 K) such as the innermost regions of protoplanetary disks \citep{Kama2019}.  This is the case of FeS with a sublimation temperature of $\sim$ 655~K \citep{Lodders2003} and the S$_n$ chains with n$>$2, which have sublimation temperatures of more than a few hundred kelvin \citep{ JimenezEscobar2011, Cazaux2022}. In the ionized gas of the M~42 nebula, dust temperatures
are high enough to sublimate this (semi-)refractory material and one would expect that sulfur is essentially undepleted. However, this is not the case for the molecular gas in the Orion Bar.  The grains located at  A$^{IF}_{\rm V}$ $>$ 3 mag in DF3 have temperatures of $\sim$ 40 $-$ 100 K, clearly insufficient to destroy refractories. One could argue that the  physical conditions in the Orion Bar are not adequate to form these compounds. \citet{Cazaux2022} proposed that sulfur allotropes form in the diffuse envelopes of molecular clouds. Based on the treatment developed by \citet{Umebayashi1980} and \citet{Draine1987} for collisions with charged grains, \citet{Ruffle1999} proposed that the sticking coefficient of positive ions such as S$^+$ increases in regions where the grains are negatively charged, hence enhancing the relative abundance of S respect to H on grain surfaces and promoting the formation of sulfur chains. This mechanism works in diffuse clouds but can be inefficient for the conditions of the Orion Bar where grain charges are more positive and dust temperatures are higher.

However, the above stationary description may not be realistic enough to simulate the expansion of an HII region into the parent molecular cloud. In OMC~1, for instance, the UV radiation from the Trapezium cluster is eroding the molecular cloud from which the stars were formed. Since sulfur refractories would survive under the physical conditions prevailing in the PDR, the lack of sulfur depletion in the Orion Bar could be the consequence of the absence of these compounds in the initial molecular cloud. 
Within the  European Millimeter Radio Astronomy Institute (IRAM) Large Program "Gas-phase Elemental abundances in Molecular cloudS" (GEMS; PI: Asunci\'on Fuente), \citet{Fuente2023} determined the sulfur elemental abundance in three cuts in Orion A. These cuts were selected in OMC-2, OMC-3, and OMC-4, at distances $>$ 1 pc from M~42.  These cuts were selected in quiescent regions, avoiding the location of protostars and  energetic outflows that could destroy interstellar grains. Nevertheless, they obtained that their observations were better explained assuming undepleted sulfur. This behavior is different from that observed by the same authors in the low-mass star-forming regions Taurus and Perseus, where sulfur was estimated to be depleted by a factor of $>$10.

Although some sulfur species such as CS and SO are routinely observed in the interstellar medium, there are very few estimates of sulfur depletion in other PDRs. \citet{Goicoechea2006} determined that sulfur depletion is $\sim$4$-$5 in the PDR associated with the Horsehead nebula on the basis  of CS and HCS$^+$ millimeter observations. The Horsehead nebula is the only region where a gas-phase doubly sulfuretted species, S$_2$H, has been detected \citep{Fuente2017-S2H}. The detection of this species is suggestive of the presence of larger sulfur chains. This PDR is illuminated by a low UV field, G$_0$ = 60, and presents a differentiated chemistry from the Orion Bar. One main difference is that the dust temperature is around $\sim$20$-$30 K, which is below  the sublimation temperature of most species, including atomic sulfur ($\sim$58~K, \citealp{Wakelam2017}). This would allow the development of a rich chemistry on the irradiated grain surfaces and most likely, the formation of progressively larger sulfur chains. \citet{Riviere2019} carried out a complete inventory of sulfur species using the "Horsehead Wide-band High-resolution Iram-30 m Surveys at two positions with Emir Receivers" (WHISPER; PI: Jerome Pety, \citealp{Pety2012}) data and found the abundances of sulfur species in the cold core close to the Horsehead nebula is similar to those found in dark clouds like TMC~1 (CP). 

Still, there are many open questions to understand sulfur chemistry in the interstellar medium. The formation and destruction mechanisms of sulfur allotropes are poorly known. In particular, there is no information on the behavior of long sulfur chains under UV irradiation. One could think that large allotropes such as S$_3$ and/or S$_8$ could be photo-dissociated on the grain surfaces, breaking into smaller chains that sublimate at lower dust temperatures. In this case, large sulfur chains could be destroyed in UV-irradiated environments although dust temperatures were below their sublimation temperatures.  If, alternatively, one thinks that the amount of sulfur in volatiles and refractories is preserved in the molecular gas during the expansion of the HII region, sulfur  depletion would be determined by the initial composition of the molecular cloud. The observation of sulfur species in PDRs with different physical conditions and their host molecular clouds would be useful to discern between these two scenarios. In this context, the study of neutral atomic sulfur in PDRs through the observation of the [S I] 25.249 $\mu$m fine-structure line using the JWST is a valuable tool for determining the amount of sulfur in volatiles and the mechanisms that subtract sulfur atoms from the gas phase in the interstellar medium. 

\section{Summary and conclusions}
The JWST (\citealp{Gardner2006}) ERS program $``$PDRs4All: Radiative feedback from massive stars$"$ has observed the prototypical PDR usually referred to as the Orion Bar as a template for the study of Galactic and extragalactic PDRs \citep{Berne2022}. We used the PDRs4All data to estimate the amount of sulfur in the Orion Bar. 
Our results can be summarized as follows:

\begin{itemize}
\item The high sensitivity of JWST has provided the first detection of the  [S I] 25.249 $\mu$m line in the Orion Bar. This is also the first detection in a PDR.
\item We have estimated a sulfur abundance of S/H $\sim$ 8$\times$10$^{-6}$ in the ionized gas. This implies a sulfur depletion lower than a factor of 2 in the Orion Veil.
\item Our team has upgraded the Meudon PDR code chemical network to account for the observations of neutral atomic sulfur.
\item Chemical modeling of DF3 shows that, as expected, most of the sulfur is in the form of S$^+$ in the outer layers of the PDR. The S$^+$/S transition takes place at a visual extinction, A$^{IF}_V$ $\sim$ 3 mag. For  A$^{IF}_V$ $>$ 3 mag, neutral atomic sulfur remains the main sulfur reservoir until A$^{IF}_V$ $\sim$5 mag. Beyond  $\sim$5 mag, most of the sulfur is locked in solid H$_2$S and the most abundant sulfur species in the gas phase are SO, SO$_2$, and H$_2$S.
\item New inelastic collision rates of atomic sulfur (S) with H, H$_2$, and He were used to carry out excitation and radiative transfer calculations to compute the [S I] 25.249 $\mu$m line intensities. Our predictions show that the emission of the  [S I] 25.249 $\mu$m line arises in molecular gas located at a visual extinction of $\sim$ 1 $-$ 5 mag from the dissociation front. In this region, the excitation of the  line is mainly through collisions with H$_2$
\item A detailed comparison of  our modeling with observations shows that sulfur is undepleted in the warm molecular gas associated with the Orion Bar.
\end{itemize}

The JWST data have allowed us  to probe the ionized and molecular gas associated with the Orion Bar. Our results show that all the observations can be explained with a moderate sulfur depletion, lower than a factor of $\sim$ 2, along this PDR. This is consistent with recent results that suggest that sulfur depletion is lower in massive star-forming regions than in dark clouds because of the higher incident UV field \citep{Fuente2023}.  Several scenarios are discussed to account for the lack of sulfur depletion in the Orion molecular cloud.

\begin{acknowledgements}
This work is based on observations made with the NASA/ESA/CSA \textit{James Webb} Space Telescope. The data were obtained from the Mikulski Archive for Space Telescopes at the Space Telescope Science Institute, which is operated by the Association of Universities for Research in Astronomy, Inc., under NASA contract NAS 5-03127 for JWST. These observations are associated with program \#1288.
Support for program \#1288 was provided by NASA through a grant from the Space Telescope Science Institute, which is operated by the Association of Universities for Research in Astronomy, Inc., under NASA contract NAS 5-03127.

AF thanks the Spanish MICIN for funding support from PID2019-106235GB-I00 and the European Research Council (ERC) for funding under the Advanced Grant project SUL4LIFE, grant agreement No101096293. AF also thanks project PID2022-137980NB-I00 funded by the Spanish Ministry of Science and Innovation/State Agency of Research MCIN/AEI/ 10.13039/501100011033 and by “ERDF A way of making Europe”.

EP acknowledges support from the University of Western Ontario, the Institute for Earth and Space Exploration, the Canadian Space Agency (CSA, 22JWGO1-16), and the Natural Sciences and Engineering Research Council of Canada. TO is supported by JSPS Bilateral Program, Grant Number 120219939.

JFB and PGY are supported in part by NASA APRA grant 80NSSC19K0698.

MGW was supported in part by NASA grant JWST-AR-01557.001-A. 

Work by YO and MR is carried out within the Collaborative Research Centre 956, sub-project C1, funded by the Deutsche Forschungsgemeinschaft (DFG) – project ID 184018867. 

JRG thanks the Spanish MCIN for funding support under grant PID2019-106110GB-I00
\end{acknowledgements}

\bibliographystyle{aa}
\bibliography{PDRs4All}

\begin{thebibliography}{102}
\expandafter\ifx\csname natexlab\endcsname\relax\def\natexlab#1{#1}\fi

\bibitem[{{Abel} {et~al.}(2004){Abel}, {Brogan}, {Ferland}, {O'Dell}, {Shaw},
  \& {Troland}}]{Abel2004}
{Abel}, N.~P., {Brogan}, C.~L., {Ferland}, G.~J., {et~al.} 2004, \apj, 609, 247

\bibitem[{{Abel} {et~al.}(2006){Abel}, {Ferland}, {O'Dell}, {Shaw}, \&
  {Troland}}]{Abel2006}
{Abel}, N.~P., {Ferland}, G.~J., {O'Dell}, C.~R., {Shaw}, G., \& {Troland},
  T.~H. 2006, \apj, 644, 344

\bibitem[{{Anderson} {et~al.}(2013){Anderson}, {Bergin}, {Maret}, \&
  {Wakelam}}]{Anderson2013}
{Anderson}, D.~E., {Bergin}, E.~A., {Maret}, S., \& {Wakelam}, V. 2013, \apj,
  779, 141

\bibitem[{{Argyriou} {et~al.}(2023){Argyriou}, {Glasse}, {Law}, {Labiano},
  {{\'A}lvarez-M{\'a}rquez}, {Patapis}, {Kavanagh}, {Gasman}, {Mueller},
  {Larson}, {Vandenbussche}, {Glauser}, {Royer}, {Dicken}, {Harkett},
  {Sargent}, {Engesser}, {Jones}, {Kendrew}, {Noriega-Crespo}, {Brandl},
  {Rieke}, {Wright}, {Lee}, \& {Wells}}]{Argyriou2023}
{Argyriou}, I., {Glasse}, A., {Law}, D.~R., {et~al.} 2023, \aap, 675, A111

\bibitem[{{Asplund} {et~al.}(2006){Asplund}, {Grevesse}, \& {Jacques
  Sauval}}]{Asplund2006}
{Asplund}, M., {Grevesse}, N., \& {Jacques Sauval}, A. 2006, \nphysa, 777, 1

\bibitem[{{Bally} {et~al.}(2000){Bally}, {O'Dell}, \&
  {McCaughrean}}]{Bally2000}
{Bally}, J., {O'Dell}, C.~R., \& {McCaughrean}, M.~J. 2000, \aj, 119, 2919

\bibitem[{{Bern{\'e}} {et~al.}(2022){Bern{\'e}}, {Habart}, {Peeters},
  {Abergel}, {Bergin}, {Bernard-Salas}, {Bron}, {Cami}, {Dartois}, {Fuente},
  {Goicoechea}, {Gordon}, {Okada}, {Onaka}, {Robberto}, {R{\"o}llig},
  {Tielens}, {Vicente}, {Wolfire}, {Alarc{\'o}n}, {Boersma}, {Canin}, {Chown},
  {Dicken}, {Languignon}, {Le Gal}, {Pound}, {Trahin}, {Simmer}, {Sidhu}, {Van
  De Putte}, {Cuadrado}, {Guilloteau}, {Maragkoudakis}, {Schefter}, {Schirmer},
  {Cazaux}, {Aleman}, {Allamandola}, {Auchettl}, {Baratta}, {Bejaoui}, {Bera},
  {Bilalbegovi{\'c}}, {Black}, {Boulanger}, {Bouwman}, {Brandl}, {Brechignac},
  {Br{\"u}nken}, {Burkhardt}, {Candian}, {Cernicharo}, {Chabot}, {Chakraborty},
  {Champion}, {Colgan}, {Cooke}, {Coutens}, {Cox}, {Demyk}, {Donovan Meyer},
  {Engrand}, {Foschino}, {Garc{\'\i}a-Lario}, {Gavilan}, {Gerin}, {Godard},
  {Gottlieb}, {Guillard}, {Gusdorf}, {Hartigan}, {He}, {Herbst}, {Hornekaer},
  {J{\"a}ger}, {Janot-Pacheco}, {Joblin}, {Kaufman}, {Kemper}, {Kendrew},
  {Kirsanova}, {Klaassen}, {Knight}, {Kwok}, {Labiano}, {Lai}, {Lee},
  {Lefloch}, {Le Petit}, {Li}, {Linz}, {Mackie}, {Madden}, {Mascetti},
  {McGuire}, {Merino}, {Micelotta}, {Misselt}, {Morse}, {Mulas}, {Neelamkodan},
  {Ohsawa}, {Omont}, {Paladini}, {Palumbo}, {Pathak}, {Pendleton},
  {Petrignani}, {Pino}, {Puga}, {Rangwala}, {Rapacioli}, {Ricca},
  {Roman-Duval}, {Roser}, {Roueff}, {Rouill{\'e}}, {Salama}, {Sales},
  {Sandstrom}, {Sarre}, {Sciamma-O'Brien}, {Sellgren}, {Shannon}, {Shenoy},
  {Teyssier}, {Thomas}, {Togi}, {Verstraete}, {Witt}, {Wootten}, {Ysard},
  {Zettergren}, {Zhang}, {Zhang}, \& {Zhen}}]{Berne2022}
{Bern{\'e}}, O., {Habart}, {\'E}., {Peeters}, E., {et~al.} 2022, \pasp, 134,
  054301

\bibitem[{Bern{\'e} {et~al.}(2024)Bern{\'e}, Habart, Peeters, Schroetter,
  Canin, Sidhu, Chown, Bron, Haworth, Klaassen, {et~al.}}]{berne2024far}
Bern{\'e}, O., Habart, E., Peeters, E., {et~al.} 2024, Science, 383, 988

\bibitem[{{Bern{\'e}} {et~al.}(2023){Bern{\'e}}, {Martin-Drumel}, {Schroetter},
  {Goicoechea}, {Jacovella}, {Gans}, {Dartois}, {Coudert}, {Bergin}, {Alarcon},
  {Cami}, {Roueff}, {Black}, {Asvany}, {Habart}, {Peeters}, {Canin}, {Trahin},
  {Joblin}, {Schlemmer}, {Thorwirth}, {Cernicharo}, {Gerin}, {Tielens},
  {Zannese}, {Abergel}, {Bernard-Salas}, {Boersma}, {Bron}, {Chown},
  {Cuadrado}, {Dicken}, {Elyajouri}, {Fuente}, {Gordon}, {Issa}, {Kannavou},
  {Khan}, {Lacinbala}, {Languignon}, {Le Gal}, {Maragkoudakis}, {Meshaka},
  {Okada}, {Onaka}, {Pasquini}, {Pound}, {Robberto}, {R{\"o}llig}, {Schefter},
  {Schirmer}, {Sidhu}, {Tabone}, {Van De Putte}, {Vicente}, \&
  {Wolfire}}]{Berne2023}
{Bern{\'e}}, O., {Martin-Drumel}, M.-A., {Schroetter}, I., {et~al.} 2023, \nat,
  621, 56

\bibitem[{{Bockel{\'e}e-Morvan} \& {Biver}(2017)}]{Bockelee2017}
{Bockel{\'e}e-Morvan}, D. \& {Biver}, N. 2017, Philosophical Transactions of
  the Royal Society of London Series A, 375, 20160252

\bibitem[{{Boffelli} {et~al.}(2023){Boffelli}, {Gauchet}, {Kashinski}, {Talbi},
  {Hickman}, {Chakrabarti}, {Bron}, {Orb{\'a}n}, {Mezei}, \&
  {Schneider}}]{Boffelli2023}
{Boffelli}, J., {Gauchet}, F., {Kashinski}, D.~O., {et~al.} 2023, \mnras, 522,
  2259

\bibitem[{{Booth} {et~al.}(2021){Booth}, {Walsh}, {Terwisscha van Scheltinga},
  {van Dishoeck}, {Ilee}, {Hogerheijde}, {Kama}, \& {Nomura}}]{Booth2021}
{Booth}, A.~S., {Walsh}, C., {Terwisscha van Scheltinga}, J., {et~al.} 2021,
  Nature Astronomy, 5, 684

\bibitem[{{Bregman} {et~al.}(1994){Bregman}, {Larson}, {Rank}, \&
  {Temi}}]{Bregman1994}
{Bregman}, J., {Larson}, K., {Rank}, D., \& {Temi}, P. 1994, \apj, 423, 326

\bibitem[{{Bron} {et~al.}(2018){Bron}, {Ag{\'u}ndez}, {Goicoechea}, \&
  {Cernicharo}}]{Bron2018}
{Bron}, E., {Ag{\'u}ndez}, M., {Goicoechea}, J.~R., \& {Cernicharo}, J. 2018,
  arXiv e-prints, arXiv:1801.01547

\bibitem[{{Bron} {et~al.}(2014){Bron}, {Le Bourlot}, \& {Le Petit}}]{Bron2014}
{Bron}, E., {Le Bourlot}, J., \& {Le Petit}, F. 2014, \aap, 569, A100

\bibitem[{{Bron} {et~al.}(2016){Bron}, {Le Petit}, \& {Le Bourlot}}]{Bron2016}
{Bron}, E., {Le Petit}, F., \& {Le Bourlot}, J. 2016, \aap, 588, A27

\bibitem[{{Bulut} {et~al.}(2021){Bulut}, {Roncero}, {Aguado}, {Loison},
  {Navarro-Almaida}, {Wakelam}, {Fuente}, {Roueff}, {Le Gal}, {Caselli},
  {Gerin}, {Hickson}, {Spezzano}, {Rivi{\'e}re-Marichalar}, {Alonso-Albi},
  {Bachiller}, {Jim{\'e}nez-Serra}, {Kramer}, {Tercero}, {Rodriguez-Baras},
  {Garc{\'\i}a-Burillo}, {Goicoechea}, {Trevi{\~n}o-Morales}, {Esplugues},
  {Cazaux}, {Commercon}, {Laas}, {Kirk}, {Lattanzi},
  {Mart{\'\i}n-Dom{\'e}nech}, {Mu{\~n}oz-Caro}, {Pineda}, {Ward-Thompson},
  {Tafalla}, {Marcelino}, {Malinen}, {Friesen}, {Giuliano}, {Ag{\'u}ndez}, \&
  {Hacar}}]{Bulut2021}
{Bulut}, N., {Roncero}, O., {Aguado}, A., {et~al.} 2021, \aap, 646, A5

\bibitem[{{Burton} {et~al.}(1990){Burton}, {Hollenbach}, \&
  {Tielens}}]{Burton1990}
{Burton}, M.~G., {Hollenbach}, D.~J., \& {Tielens}, A.~G.~G.~M. 1990, \apj,
  365, 620

\bibitem[{{Capria} {et~al.}(2017){Capria}, {Capaccioni}, {Filacchione}, {Tosi},
  {De Sanctis}, {Mottola}, {Ciarniello}, {Formisano}, {Longobardo},
  {Migliorini}, {Palomba}, {Raponi}, {K{\"u}hrt}, {Bockel{\'e}e-Morvan},
  {Erard}, {Leyrat}, \& {Zinzi}}]{Capria2017}
{Capria}, M.~T., {Capaccioni}, F., {Filacchione}, G., {et~al.} 2017, \mnras,
  469, S685

\bibitem[{{Cardelli} {et~al.}(1989){Cardelli}, {Clayton}, \&
  {Mathis}}]{Cardelli1989}
{Cardelli}, J.~A., {Clayton}, G.~C., \& {Mathis}, J.~S. 1989, \apj, 345, 245

\bibitem[{{Caselli}(2020)}]{Caselli2020}
{Caselli}, P. 2020, Physics of Life Reviews, 32, 117

\bibitem[{{Cazaux} {et~al.}(2022){Cazaux}, {Carrascosa}, {Mu{\~n}oz Caro},
  {Caselli}, {Fuente}, {Navarro-Almaida}, \&
  {Rivi{\'e}re-Marichalar}}]{Cazaux2022}
{Cazaux}, S., {Carrascosa}, H., {Mu{\~n}oz Caro}, G.~M., {et~al.} 2022, \aap,
  657, A100

\bibitem[{{Chown} {et~al.}(2023){Chown}, {Sidhu}, {Peeters}, {Tielens}, {Cami},
  {Bern{\'e}}, {Habart}, {Alarc{\'o}n}, {Canin}, {Schroetter}, {Trahin}, {Van
  De Putte}, {Abergel}, {Bergin}, {Bernard-Salas}, {Boersma}, {Bron},
  {Cuadrado}, {Dartois}, {Dicken}, {El-Yajouri}, {Fuente}, {Goicoechea},
  {Gordon}, {Issa}, {Joblin}, {Kannavou}, {Khan}, {Lacinbala}, {Languignon},
  {Le Gal}, {Maragkoudakis}, {Meshaka}, {Okada}, {Onaka}, {Pasquini}, {Pound},
  {Robberto}, {R{\"o}llig}, {Schefter}, {Schirmer}, {Vicente}, {Wolfire},
  {Zannese}, {Aleman}, {Allamandola}, {Auchettl}, {Baratta}, {Bejaoui}, {Bera},
  {Black}, {Boulanger}, {Bouwman}, {Brandl}, {Brechignac}, {Br{\"u}nken},
  {Buragohain}, {Burkhardt}, {Candian}, {Cazaux}, {Cernicharo}, {Chabot},
  {Chakraborty}, {Champion}, {Colgan}, {Cooke}, {Coutens}, {Cox}, {Demyk},
  {Donovan Meyer}, {Foschino}, {Garc{\'\i}a-Lario}, {Gavilan}, {Gerin},
  {Gottlieb}, {Guillard}, {Gusdorf}, {Hartigan}, {He}, {Herbst}, {Hornekaer},
  {J{\"a}ger}, {Janot-Pacheco}, {Kaufman}, {Kemper}, {Kendrew}, {Kirsanova},
  {Klaassen}, {Kwok}, {Labiano}, {Lai}, {Lee}, {Lefloch}, {Le Petit}, {Li},
  {Linz}, {Mackie}, {Madden}, {Mascetti}, {McGuire}, {Merino}, {Micelotta},
  {Misselt}, {Morse}, {Mulas}, {Neelamkodan}, {Ohsawa}, {Omont}, {Paladini},
  {Palumbo}, {Pathak}, {Pendleton}, {Petrignani}, {Pino}, {Puga}, {Rangwala},
  {Rapacioli}, {Ricca}, {Roman-Duval}, {Roser}, {Roueff}, {Rouill{\'e}},
  {Salama}, {Sales}, {Sandstrom}, {Sarre}, {Sciamma-O'Brien}, {Sellgren},
  {Shenoy}, {Teyssier}, {Thomas}, {Togi}, {Verstraete}, {Witt}, {Wootten},
  {Zettergren}, {Zhang}, {Zhang}, \& {Zhen}}]{Chown2023}
{Chown}, R., {Sidhu}, A., {Peeters}, E., {et~al.} 2023, arXiv e-prints,
  arXiv:2308.16733

\bibitem[{{Daflon} {et~al.}(2009){Daflon}, {Cunha}, {de la Reza}, {Holtzman},
  \& {Chiappini}}]{Daflon2009}
{Daflon}, S., {Cunha}, K., {de la Reza}, R., {Holtzman}, J., \& {Chiappini}, C.
  2009, \aj, 138, 1577

\bibitem[{{Das} {et~al.}(2018){Das}, {Sil}, {Gorai}, {Chakrabarti}, \&
  {Loison}}]{Das2018}
{Das}, A., {Sil}, M., {Gorai}, P., {Chakrabarti}, S.~K., \& {Loison}, J.~C.
  2018, \apjs, 237, 9

\bibitem[{{Del Zanna} {et~al.}(2021){Del Zanna}, {Dere}, {Young}, \&
  {Landi}}]{DelZanna2021}
{Del Zanna}, G., {Dere}, K.~P., {Young}, P.~R., \& {Landi}, E. 2021, \apj, 909,
  38

\bibitem[{{Draine} \& {Sutin}(1987)}]{Draine1987}
{Draine}, B.~T. \& {Sutin}, B. 1987, \apj, 320, 803

\bibitem[{{Ehrenfreund} {et~al.}(2015){Ehrenfreund}, {Elsaesser}, \&
  {Groen}}]{Ehrenfreund2015}
{Ehrenfreund}, P., {Elsaesser}, A., \& {Groen}, J. 2015, Highlights of
  Astronomy, 16, 709

\bibitem[{{Esteban} {et~al.}(2004){Esteban}, {Peimbert}, {Garc{\'\i}a-Rojas},
  {Ruiz}, {Peimbert}, \& {Rodr{\'\i}guez}}]{Esteban2004}
{Esteban}, C., {Peimbert}, M., {Garc{\'\i}a-Rojas}, J., {et~al.} 2004, \mnras,
  355, 229

\bibitem[{{Flower}(1990)}]{Flower1990}
{Flower}, D.~R. 1990, \mnras, 242, 1P

\bibitem[{{Fuente} {et~al.}(2017{\natexlab{a}}){Fuente}, {Gerin}, {Pety},
  {Commer{\c{c}}on}, {Ag{\'u}ndez}, {Cernicharo}, {Marcelino}, {Roueff}, {Lis},
  \& {Wootten}}]{Fuente2017-B1}
{Fuente}, A., {Gerin}, M., {Pety}, J., {et~al.} 2017{\natexlab{a}}, \aap, 606,
  L3

\bibitem[{{Fuente} {et~al.}(2017{\natexlab{b}}){Fuente}, {Goicoechea}, {Pety},
  {Le Gal}, {Mart{\'\i}n-Dom{\'e}nech}, {Gratier}, {Guzm{\'a}n}, {Roueff},
  {Loison}, {Mu{\~n}oz Caro}, {Wakelam}, {Gerin}, {Riviere-Marichalar}, \&
  {Vidal}}]{Fuente2017-S2H}
{Fuente}, A., {Goicoechea}, J.~R., {Pety}, J., {et~al.} 2017{\natexlab{b}},
  \apjl, 851, L49

\bibitem[{{Fuente} {et~al.}(2019){Fuente}, {Navarro}, {Caselli}, {Gerin},
  {Kramer}, {Roueff}, {Alonso-Albi}, {Bachiller}, {Cazaux}, {Commercon},
  {Friesen}, {Garc{\'\i}a-Burillo}, {Giuliano}, {Goicoechea}, {Gratier},
  {Hacar}, {Jim{\'e}nez-Serra}, {Kirk}, {Lattanzi}, {Loison}, {Malinen},
  {Marcelino}, {Mart{\'\i}n-Dom{\'e}nech}, {Mu{\~n}oz-Caro}, {Pineda},
  {Tafalla}, {Tercero}, {Ward-Thompson}, {Trevi{\~n}o-Morales},
  {Rivi{\'e}re-Marichalar}, {Roncero}, {Vidal}, \& {Ballester}}]{Fuente2019}
{Fuente}, A., {Navarro}, D.~G., {Caselli}, P., {et~al.} 2019, \aap, 624, A105

\bibitem[{{Fuente} {et~al.}(2023){Fuente}, {Rivi{\`e}re-Marichalar},
  {Beitia-Antero}, {Caselli}, {Wakelam}, {Esplugues}, {Rodr{\'\i}guez-Baras},
  {Navarro-Almaida}, {Gerin}, {Kramer}, {Bachiller}, {Goicoechea},
  {Jim{\'e}nez-Serra}, {Loison}, {Ivlev}, {Mart{\'\i}n-Dom{\'e}nech},
  {Spezzano}, {Roncero}, {Mu{\~n}oz-Caro}, {Cazaux}, \&
  {Marcelino}}]{Fuente2023}
{Fuente}, A., {Rivi{\`e}re-Marichalar}, P., {Beitia-Antero}, L., {et~al.} 2023,
  \aap, 670, A114

\bibitem[{{Fuente} {et~al.}(2003){Fuente}, {Rodr{\i}guez-Franco},
  {Garc{\i}a-Burillo}, {Mart{\i}n-Pintado}, \& {Black}}]{Fuente2003}
{Fuente}, A., {Rodr{\i}guez-Franco}, A., {Garc{\i}a-Burillo}, S.,
  {Mart{\i}n-Pintado}, J., \& {Black}, J.~H. 2003, \aap, 406, 899

\bibitem[{{Fuente} {et~al.}(1996){Fuente}, {Rodriguez-Franco}, \&
  {Martin-Pintado}}]{Fuente1996}
{Fuente}, A., {Rodriguez-Franco}, A., \& {Martin-Pintado}, J. 1996, \aap, 312,
  599

\bibitem[{{Gardner} {et~al.}(2006){Gardner}, {Mather}, {Clampin}, {Doyon},
  {Greenhouse}, {Hammel}, {Hutchings}, {Jakobsen}, {Lilly}, {Long}, {Lunine},
  {McCaughrean}, {Mountain}, {Nella}, {Rieke}, {Rieke}, {Rix}, {Smith},
  {Sonneborn}, {Stiavelli}, {Stockman}, {Windhorst}, \& {Wright}}]{Gardner2006}
{Gardner}, J.~P., {Mather}, J.~C., {Clampin}, M., {et~al.} 2006, \ssr, 123, 485

\bibitem[{{Goicoechea} {et~al.}(2021){Goicoechea}, {Aguado}, {Cuadrado},
  {Roncero}, {Pety}, {Bron}, {Fuente}, {Riquelme}, {Chapillon}, {Herrera}, \&
  {Duran}}]{Goicoechea2021}
{Goicoechea}, J.~R., {Aguado}, A., {Cuadrado}, S., {et~al.} 2021, \aap, 647,
  A10

\bibitem[{{Goicoechea} \& {Cuadrado}(2021)}]{GoicoecheaCuadrado2021}
{Goicoechea}, J.~R. \& {Cuadrado}, S. 2021, \aap, 647, L7

\bibitem[{{Goicoechea} {et~al.}(2011){Goicoechea}, {Joblin}, {Contursi},
  {Bern{\'e}}, {Cernicharo}, {Gerin}, {Le Bourlot}, {Bergin}, {Bell}, \&
  {R{\"o}llig}}]{Goicoechea2011}
{Goicoechea}, J.~R., {Joblin}, C., {Contursi}, A., {et~al.} 2011, \aap, 530,
  L16

\bibitem[{{Goicoechea} \& {Le Bourlot}(2007)}]{Goicoechea2007}
{Goicoechea}, J.~R. \& {Le Bourlot}, J. 2007, \aap, 467, 1

\bibitem[{{Goicoechea} {et~al.}(2016){Goicoechea}, {Pety}, {Cuadrado},
  {Cernicharo}, {Chapillon}, {Fuente}, {Gerin}, {Joblin}, {Marcelino}, \&
  {Pilleri}}]{Goicoechea2016}
{Goicoechea}, J.~R., {Pety}, J., {Cuadrado}, S., {et~al.} 2016, \nat, 537, 207

\bibitem[{{Goicoechea} {et~al.}(2006){Goicoechea}, {Pety}, {Gerin}, {Teyssier},
  {Roueff}, {Hily-Blant}, \& {Baek}}]{Goicoechea2006}
{Goicoechea}, J.~R., {Pety}, J., {Gerin}, M., {et~al.} 2006, \aap, 456, 565

\bibitem[{{Gonzalez Garcia} {et~al.}(2008){Gonzalez Garcia}, {Le Bourlot}, {Le
  Petit}, \& {Roueff}}]{Gonzalez2008}
{Gonzalez Garcia}, M., {Le Bourlot}, J., {Le Petit}, F., \& {Roueff}, E. 2008,
  \aap, 485, 127

\bibitem[{{Gordon} {et~al.}(2022){Gordon}, {Bohlin}, {Sloan}, {Rieke}, {Volk},
  {Boyer}, {Muzerolle}, {Schlawin}, {Deustua}, {Hines}, {Kraemer}, {Mullally},
  \& {Su}}]{Gordon2022}
{Gordon}, K.~D., {Bohlin}, R., {Sloan}, G.~C., {et~al.} 2022, \aj, 163, 267

\bibitem[{{Guzm{\'a}n} {et~al.}(2012){Guzm{\'a}n}, {Pety}, {Gratier},
  {Goicoechea}, {Gerin}, {Roueff}, \& {Teyssier}}]{Pety2012}
{Guzm{\'a}n}, V., {Pety}, J., {Gratier}, P., {et~al.} 2012, \aap, 543, L1

\bibitem[{{Guzm{\'a}n} {et~al.}(2017){Guzm{\'a}n}, {{\"O}berg}, {Huang},
  {Loomis}, \& {Qi}}]{Guzman2017}
{Guzm{\'a}n}, V.~V., {{\"O}berg}, K.~I., {Huang}, J., {Loomis}, R., \& {Qi}, C.
  2017, \apj, 836, 30

\bibitem[{{Haas} {et~al.}(1986){Haas}, {Hollenbach}, \& {Erickson}}]{Haas1986}
{Haas}, M.~R., {Hollenbach}, D.~J., \& {Erickson}, E.~F. 1986, \apjl, 301, L57

\bibitem[{{Habart} {et~al.}(2023){Habart}, {Peeters}, {Bern{\'e}}, {Trahin},
  {Canin}, {Chown}, {Sidhu}, {Van De Putte}, {Alarc{\'o}n}, {Schroetter},
  {Dartois}, {Vicente}, {Abergel}, {Bergin}, {Bernard-Salas}, {Boersma},
  {Bron}, {Cami}, {Cuadrado}, {Dicken}, {Elyajouri}, {Fuente}, {Goicoechea},
  {Gordon}, {Issa}, {Joblin}, {Kannavou}, {Khan}, {Lacinbala}, {Languignon},
  {Le Gal}, {Maragkoudakis}, {Meshaka}, {Okada}, {Onaka}, {Pasquini}, {Pound},
  {Robberto}, {R{\"o}llig}, {Schefter}, {Schirmer}, {Tabone}, {Tielens},
  {Wolfire}, {Zannese}, {Ysard}, {Miville-Deschenes}, {Aleman}, {Allamandola},
  {Auchettl}, {Baratta}, {Bejaoui}, {Bera}, {Black}, {Boulanger}, {Bouwman},
  {Brandl}, {Brechignac}, {Br{\"u}nken}, {Buragohain}, {Burkhardt}, {Candian},
  {Cazaux}, {Cernicharo}, {Chabot}, {Chakraborty}, {Champion}, {Colgan},
  {Cooke}, {Coutens}, {Cox}, {Demyk}, {Donovan Meyer}, {Foschino},
  {Garc{\'\i}a-Lario}, {Gavilan}, {Gerin}, {Gottlieb}, {Guillard}, {Gusdorf},
  {Hartigan}, {He}, {Herbst}, {Hornekaer}, {J{\"a}ger}, {Janot-Pacheco},
  {Kaufman}, {Kemper}, {Kendrew}, {Kirsanova}, {Klaassen}, {Kwok}, {Labiano},
  {Lai}, {Lee}, {Lefloch}, {Le Petit}, {Li}, {Linz}, {Mackie}, {Madden},
  {Mascetti}, {McGuire}, {Merino}, {Micelotta}, {Misselt}, {Morse}, {Mulas},
  {Neelamkodan}, {Ohsawa}, {Omont}, {Paladini}, {Palumbo}, {Pathak},
  {Pendleton}, {Petrignani}, {Pino}, {Puga}, {Rangwala}, {Rapacioli}, {Ricca},
  {Roman-Duval}, {Roser}, {Roueff}, {Rouill{\'e}}, {Salama}, {Sales},
  {Sandstrom}, {Sarre}, {Sciamma-O'Brien}, {Sellgren}, {Shenoy}, {Teyssier},
  {Thomas}, {Togi}, {Verstraete}, {Witt}, {Wootten}, {Zettergren}, {Zhang},
  {Zhang}, \& {Zhen}}]{Habart2023}
{Habart}, E., {Peeters}, E., {Bern{\'e}}, O., {et~al.} 2023, arXiv e-prints,
  arXiv:2308.16732

\bibitem[{{Hasegawa} \& {Herbst}(1993)}]{HasegawaHerbst1993}
{Hasegawa}, T.~I. \& {Herbst}, E. 1993, \mnras, 261, 83

\bibitem[{{Hayashi} {et~al.}(1985){Hayashi}, {Hasegawa}, {Gatley}, {Garden}, \&
  {Kaifu}}]{Hayashi1985}
{Hayashi}, M., {Hasegawa}, T., {Gatley}, I., {Garden}, R., \& {Kaifu}, N. 1985,
  \mnras, 215, 31P

\bibitem[{{Heays} {et~al.}(2017){Heays}, {Bosman}, \& {van
  Dishoeck}}]{Heays2017}
{Heays}, A.~N., {Bosman}, A.~D., \& {van Dishoeck}, E.~F. 2017, \aap, 602, A105

\bibitem[{{Hily-Blant} {et~al.}(2022){Hily-Blant}, {Pineau des For{\^e}ts},
  {Faure}, \& {Lique}}]{HilyBlant2022}
{Hily-Blant}, P., {Pineau des For{\^e}ts}, G., {Faure}, A., \& {Lique}, F.
  2022, \aap, 658, A168

\bibitem[{{Hogerheijde} {et~al.}(1995){Hogerheijde}, {Jansen}, \& {van
  Dishoeck}}]{Hogerheijde1995}
{Hogerheijde}, M.~R., {Jansen}, D.~J., \& {van Dishoeck}, E.~F. 1995, \aap,
  294, 792

\bibitem[{{Hrodmarsson} \& {van Dishoeck}(2023)}]{Hrodmarsson2023}
{Hrodmarsson}, H.~R. \& {van Dishoeck}, E.~F. 2023, \aap, 675, A25

\bibitem[{{Jaquet} {et~al.}(1992){Jaquet}, {Staemmler}, {Smith}, \&
  {Flower}}]{Jaquet1992}
{Jaquet}, R., {Staemmler}, V., {Smith}, M.~D., \& {Flower}, D.~R. 1992, J.
  Phys. B At. Molec. Phys., 25, 285

\bibitem[{{Jensen} {et~al.}(2019){Jensen}, {J{\o}rgensen}, {Kristensen},
  {Furuya}, {Coutens}, {van Dishoeck}, {Harsono}, \& {Persson}}]{Jensen2019}
{Jensen}, S.~S., {J{\o}rgensen}, J.~K., {Kristensen}, L.~E., {et~al.} 2019,
  \aap, 631, A25

\bibitem[{{Jim{\'e}nez-Escobar} \& {Mu{\~n}oz Caro}(2011)}]{JimenezEscobar2011}
{Jim{\'e}nez-Escobar}, A. \& {Mu{\~n}oz Caro}, G.~M. 2011, \aap, 536, A91

\bibitem[{{Jim{\'e}nez-Escobar} {et~al.}(2012){Jim{\'e}nez-Escobar}, {Mu{\~n}oz
  Caro}, {Ciaravella}, {Cecchi-Pestellini}, {Candia}, \&
  {Micela}}]{JimenezEscobar2012}
{Jim{\'e}nez-Escobar}, A., {Mu{\~n}oz Caro}, G.~M., {Ciaravella}, A., {et~al.}
  2012, \apjl, 751, L40

\bibitem[{{Joblin} {et~al.}(2018){Joblin}, {Bron}, {Pinto}, {Pilleri}, {Le
  Petit}, {Gerin}, {Le Bourlot}, {Fuente}, {Berne}, {Goicoechea}, {Habart},
  {K{\"o}hler}, {Teyssier}, {Nagy}, {Montillaud}, {Vastel}, {Cernicharo},
  {R{\"o}llig}, {Ossenkopf-Okada}, \& {Bergin}}]{Joblin2018}
{Joblin}, C., {Bron}, E., {Pinto}, C., {et~al.} 2018, \aap, 615, A129

\bibitem[{{Kama} {et~al.}(2019){Kama}, {Shorttle}, {Jermyn}, {Folsom},
  {Furuya}, {Bergin}, {Walsh}, \& {Keller}}]{Kama2019}
{Kama}, M., {Shorttle}, O., {Jermyn}, A.~S., {et~al.} 2019, \apj, 885, 114

\bibitem[{Kramida {et~al.}(2022)Kramida, {Yu.~Ralchenko}, Reader, \& {and NIST
  ASD Team}}]{NIST_ASD}
Kramida, A., {Yu.~Ralchenko}, Reader, J., \& {and NIST ASD Team}. 2022, {NIST
  Atomic Spectra Database (ver. 5.10), [Online]. Available:
  {\tt{https://physics.nist.gov/asd}} [2023, November 20]. National Institute
  of Standards and Technology, Gaithersburg, MD.}

\bibitem[{{Le Bourlot} {et~al.}(2012){Le Bourlot}, {Le Petit}, {Pinto},
  {Roueff}, \& {Roy}}]{Bourlot2012}
{Le Bourlot}, J., {Le Petit}, F., {Pinto}, C., {Roueff}, E., \& {Roy}, F. 2012,
  \aap, 541, A76

\bibitem[{{Lique} {et~al.}(2018){Lique}, {K{\l}os}, \& {Le Picard}}]{Lique2018}
{Lique}, F., {K{\l}os}, J., \& {Le Picard}, S.~D. 2018, Physical Chemistry
  Chemical Physics (Incorporating Faraday Transactions), 20, 5427

\bibitem[{{Lodders}(2003)}]{Lodders2003}
{Lodders}, K. 2003, \apj, 591, 1220

\bibitem[{{Marconi} {et~al.}(1998){Marconi}, {Testi}, {Natta}, \&
  {Walmsley}}]{Marconi1998}
{Marconi}, A., {Testi}, L., {Natta}, A., \& {Walmsley}, C.~M. 1998, \aap, 330,
  696

\bibitem[{{Mathis} {et~al.}(1977){Mathis}, {Rumpl}, \&
  {Nordsieck}}]{Mathis1977}
{Mathis}, J.~S., {Rumpl}, W., \& {Nordsieck}, K.~H. 1977, \apj, 217, 425

\bibitem[{{McLeod} {et~al.}(2016){McLeod}, {Weilbacher}, {Ginsburg}, {Dale},
  {Ramsay}, \& {Testi}}]{McLeod2016}
{McLeod}, A.~F., {Weilbacher}, P.~M., {Ginsburg}, A., {et~al.} 2016, \mnras,
  455, 4057

\bibitem[{{Menten} {et~al.}(2007){Menten}, {Reid}, {Forbrich}, \&
  {Brunthaler}}]{Menten2007}
{Menten}, K.~M., {Reid}, M.~J., {Forbrich}, J., \& {Brunthaler}, A. 2007, \aap,
  474, 515

\bibitem[{{Meyer} {et~al.}(1997){Meyer}, {Cardelli}, \& {Sofia}}]{Meyer1997}
{Meyer}, D.~M., {Cardelli}, J.~A., \& {Sofia}, U.~J. 1997, \apjl, 490, L103

\bibitem[{{Meyer} {et~al.}(1998){Meyer}, {Jura}, \& {Cardelli}}]{Meyer1998}
{Meyer}, D.~M., {Jura}, M., \& {Cardelli}, J.~A. 1998, \apj, 493, 222

\bibitem[{{Neufeld} {et~al.}(2015){Neufeld}, {Godard}, {Gerin}, {Pineau des
  For{\^e}ts}, {Bernier}, {Falgarone}, {Graf}, {G{\"u}sten}, {Herbst},
  {Lesaffre}, {Schilke}, {Sonnentrucker}, \& {Wiesemeyer}}]{Neufeld2015}
{Neufeld}, D.~A., {Godard}, B., {Gerin}, M., {et~al.} 2015, \aap, 577, A49

\bibitem[{{Omodaka} {et~al.}(1994){Omodaka}, {Hayashi}, {Hasegawa}, \&
  {Hayashi}}]{Omodaka1994}
{Omodaka}, T., {Hayashi}, M., {Hasegawa}, T., \& {Hayashi}, S.~S. 1994, \apj,
  430, 256

\bibitem[{{Ossenkopf} \& {Henning}(1994)}]{Ossenkopf1994}
{Ossenkopf}, V. \& {Henning}, T. 1994, \aap, 291, 943

\bibitem[{{Patapis} {et~al.}(2024){Patapis}, {Argyriou}, {Law}, {Glauser},
  {Glasse}, {Labiano}, {{\'A}lvarez-M{\'a}rquez}, {Kavanagh}, {Gasman},
  {Mueller}, {Larson}, {Vandenbussche}, {Lee}, {Klaassen}, {Guillard}, \&
  {Wright}}]{Patapis2024}
{Patapis}, P., {Argyriou}, I., {Law}, D.~R., {et~al.} 2024, \aap, 682, A53

\bibitem[{{Peeters} {et~al.}(2023){Peeters}, {Habart}, {Berne}, {Sidhu},
  {Chown}, {Van De Putte}, {Trahin}, {Schroetter}, {Canin}, {Alarcon},
  {Schefter}, {Khan}, {Pasquini}, {Tielens}, {Wolfire}, {Dartois},
  {Goicoechea}, {Maragkoudakis}, {Onaka}, {Pound}, {Vicente}, {Abergel},
  {Bergin}, {Bernard-Salas}, {Boersma}, {Bron}, {Cami}, {Cuadrado}, {Dicken},
  {Elyajour}, {Fuente}, {Gordon}, {Issa}, {Joblin}, {Kannavou}, {Lacinbala},
  {Languignon}, {Le Gal}, {Meshaka}, {Okada}, {Robberto}, {Roellig},
  {Schirmer}, {Tabone}, {Zannese}, {Aleman}, {Allamandola}, {Auchettl},
  {Baratta}, {Bejaoui}, {Bera}, {Black}, {Boulanger}, {Bouwman}, {Brandl},
  {Brechignac}, {Brunken}, {Buragohain}, {Burkhardt}, {Candian}, {Cazaux},
  {Cernicharo}, {Chabot}, {Chakraborty}, {Champion}, {Colgan}, {Cooke},
  {Coutens}, {Cox}, {Demyk}, {Donovan Meyer}, {Foschino}, {Garcia-Lario},
  {Gerin}, {Gottlieb}, {Guillard}, {Gusdorf}, {Hartigan}, {He}, {Herbst},
  {Hornekaer}, {Jager}, {Janot-Pacheco}, {Kaufman}, {Kendrew}, {Kirsanova},
  {Klaassen}, {Kwok}, {Labiano}, {Lai}, {Lee}, {Lefloch}, {Le Petit}, {Li},
  {Linz}, {Mackie}, {Madden}, {Mascetti}, {McGuire}, {Merino}, {Micelotta},
  {Misselt}, {Morse}, {Mulas}, {Neelamkodan}, {Ohsawa}, {Paladini}, {Palumbo},
  {Pathak}, {Pendleton}, {Petrignani}, {Pino}, {Puga}, {Rangwala}, {Rapacioli},
  {Ricca}, {Roman-Duval}, {Roser}, {Roueff}, {Rouille}, {Salama}, {Sales},
  {Sandstrom}, {Sarre}, {Sciamma-O'Brien}, {Sellgren}, {Shenoy}, {Teyssier},
  {Thomas}, {Togi}, {Verstraete}, {Witt}, {Wootten}, {Ysard}, {Zettergren},
  {Zhang}, {Zhang}, \& {Zhen}}]{Peeters2023}
{Peeters}, E., {Habart}, E., {Berne}, O., {et~al.} 2023, arXiv e-prints,
  arXiv:2310.08720

\bibitem[{{Peimbert} \& {Torres-Peimbert}(1977)}]{Peimbert1977}
{Peimbert}, M. \& {Torres-Peimbert}, S. 1977, \mnras, 179, 217

\bibitem[{{Perrero} {et~al.}(2024){Perrero}, {Beitia-Antero}, {Fuente},
  {Ugliengo}, \& {Rimola}}]{Perrero2024}
{Perrero}, J., {Beitia-Antero}, L., {Fuente}, A., {Ugliengo}, P., \& {Rimola},
  A. 2024, \mnras, 527, 10697

\bibitem[{{Perrero} {et~al.}(2022){Perrero}, {Enrique-Romero}, {Ferrero},
  {Ceccarelli}, {Podio}, {Codella}, {Rimola}, \& {Ugliengo}}]{Perrero2022}
{Perrero}, J., {Enrique-Romero}, J., {Ferrero}, S., {et~al.} 2022, \apj, 938,
  158

\bibitem[{{Rivi{\`e}re-Marichalar} {et~al.}(2019){Rivi{\`e}re-Marichalar},
  {Fuente}, {Goicoechea}, {Pety}, {Le Gal}, {Gratier}, {Guzm{\'a}n}, {Roueff},
  {Loison}, {Wakelam}, \& {Gerin}}]{Riviere2019}
{Rivi{\`e}re-Marichalar}, P., {Fuente}, A., {Goicoechea}, J.~R., {et~al.} 2019,
  \aap, 628, A16

\bibitem[{{Roueff} {et~al.}(2019){Roueff}, {Abgrall}, {Czachorowski},
  {Pachucki}, {Puchalski}, \& {Komasa}}]{Roueff2019}
{Roueff}, E., {Abgrall}, H., {Czachorowski}, P., {et~al.} 2019, \aap, 630, A58

\bibitem[{{Rubin} {et~al.}(2011){Rubin}, {Simpson}, {O'Dell}, {McNabb},
  {Colgan}, {Zhuge}, {Ferland}, \& {Hidalgo}}]{Rubin2011}
{Rubin}, R.~H., {Simpson}, J.~P., {O'Dell}, C.~R., {et~al.} 2011, \mnras, 410,
  1320

\bibitem[{{Ruffle} {et~al.}(1999){Ruffle}, {Hartquist}, {Caselli}, \&
  {Williams}}]{Ruffle1999}
{Ruffle}, D.~P., {Hartquist}, T.~W., {Caselli}, P., \& {Williams}, D.~A. 1999,
  \mnras, 306, 691

\bibitem[{{Savage} \& {Sembach}(1996)}]{Savage1996}
{Savage}, B.~D. \& {Sembach}, K.~R. 1996, \araa, 34, 279

\bibitem[{{Shingledecker} {et~al.}(2020){Shingledecker}, {Lamberts}, {Laas},
  {Vasyunin}, {Herbst}, {K{\"a}stner}, \& {Caselli}}]{Shingledecker2020}
{Shingledecker}, C.~N., {Lamberts}, T., {Laas}, J.~C., {et~al.} 2020, \apj,
  888, 52

\bibitem[{{Sofia} {et~al.}(1994){Sofia}, {Cardelli}, \& {Savage}}]{Sofia1994}
{Sofia}, U.~J., {Cardelli}, J.~A., \& {Savage}, B.~D. 1994, \apj, 430, 650

\bibitem[{{Storey} \& {Hummer}(1995)}]{Storey1995}
{Storey}, P.~J. \& {Hummer}, D.~G. 1995, \mnras, 272, 41

\bibitem[{{Tauber} {et~al.}(1995){Tauber}, {Lis}, {Keene}, {Schilke}, \&
  {Buettgenbach}}]{Tauber1995}
{Tauber}, J.~A., {Lis}, D.~C., {Keene}, J., {Schilke}, P., \& {Buettgenbach},
  T.~H. 1995, \aap, 297, 567

\bibitem[{{Tauber} {et~al.}(1994){Tauber}, {Tielens}, {Meixner}, \&
  {Goldsmith}}]{Tauber1994}
{Tauber}, J.~A., {Tielens}, A.~G.~G.~M., {Meixner}, M., \& {Goldsmith}, P.~F.
  1994, \apj, 422, 136

\bibitem[{{Tayal}(2004)}]{Tayal2004}
{Tayal}, S.~S. 2004, \apjs, 153, 581

\bibitem[{{Umebayashi} \& {Nakano}(1980)}]{Umebayashi1980}
{Umebayashi}, T. \& {Nakano}, T. 1980, \pasj, 32, 405

\bibitem[{{Van De Putte} {et~al.}(2024){Van De Putte}, {Meshaka}, {Trahin},
  {Habart}, {Peeters}, {Bern{\'e}}, {Alarc{\'o}n}, {Canin}, {Chown},
  {Schroetter}, {Sidhu}, {Boersma}, {Bron}, {Dartois}, {Goicoechea}, {Gordon},
  {Onaka}, {Tielens}, {Verstraete}, {Wolfire}, {Abergel}, {Bergin},
  {Bernard-Salas}, {Cami}, {Cuadrado}, {Dicken}, {Elyajouri}, {Fuente},
  {Joblin}, {Khan}, {Lacinbala}, {Languignon}, {Le Gal}, {Maragkoudakis},
  {Okada}, {Pasquini}, {Pound}, {Robberto}, {R{\"o}llig}, {Schefter},
  {Schirmer}, {Tabone}, {Vicente}, {Zannese}, {Colgan}, {He}, {Rouill{\'e}},
  {Togi}, {Aleman}, {Auchettl}, {Baratta}, {Bejaoui}, {Bera}, {Black},
  {Boulanger}, {Bouwman}, {Brandl}, {Brechignac}, {Br{\"u}nken}, {Buragohain},
  {Burkhardt}, {Candian}, {Cazaux}, {Cernicharo}, {Chabot}, {Chakraborty},
  {Champion}, {Cooke}, {Coutens}, {Cox}, {Demyk}, {Donovan Meyer}, {Foschino},
  {Garc{\'\i}a-Lario}, {Gerin}, {Gottlieb}, {Guillard}, {Gusdorf}, {Hartigan},
  {Herbst}, {Hornekaer}, {Issa}, {J{\"a}ger}, {Janot-Pacheco}, {Kannavou},
  {Kaufman}, {Kemper}, {Kendrew}, {Kirsanova}, {Klaassen}, {Kwok}, {Labiano},
  {Lai}, {Le Floch}, {Le Petit}, {Li}, {Linz}, {Mackie}, {Madden}, {Mascetti},
  {McGuire}, {Merino}, {Micelotta}, {Morse}, {Mulas}, {Neelamkodan}, {Ohsawa},
  {Omont}, {Paladini}, {Palumbo}, {Pathak}, {Pendleton}, {Petrignani}, {Pino},
  {Puga}, {Rangwala}, {Rapacioli}, {Rho}, {Ricca}, {Roman-Duval}, {Roser},
  {Roueff}, {Salama}, {Sales}, {Sandstrom}, {Sarre}, {Sciamma-O'Brien},
  {Sellgren}, {Shenoy}, {Teyssier}, {Thomas}, {Witt}, {Wootten}, {Ysard},
  {Zettergren}, {Zhang}, {Zhang}, \& {Zhen}}]{Putte2024}
{Van De Putte}, D., {Meshaka}, R., {Trahin}, B., {et~al.} 2024, arXiv e-prints,
  arXiv:2404.03111

\bibitem[{{Vastel} {et~al.}(2018){Vastel}, {Qu{\'e}nard}, {Le Gal}, {Wakelam},
  {Andrianasolo}, {Caselli}, {Vidal}, {Ceccarelli}, {Lefloch}, \&
  {Bachiller}}]{Vastel2018}
{Vastel}, C., {Qu{\'e}nard}, D., {Le Gal}, R., {et~al.} 2018, \mnras, 478, 5514

\bibitem[{{Vidal} {et~al.}(2017){Vidal}, {Loison}, {Jaziri}, {Ruaud},
  {Gratier}, \& {Wakelam}}]{Vidal2017-DC}
{Vidal}, T.~H.~G., {Loison}, J.-C., {Jaziri}, A.~Y., {et~al.} 2017, \mnras,
  469, 435

\bibitem[{{Wakelam} {et~al.}(2017){Wakelam}, {Loison}, {Mereau}, \&
  {Ruaud}}]{Wakelam2017}
{Wakelam}, V., {Loison}, J.~C., {Mereau}, R., \& {Ruaud}, M. 2017, Molecular
  Astrophysics, 6, 22

\bibitem[{{Wang} {et~al.}(2013){Wang}, {Bourke}, {Hogerheijde}, {van der Tak},
  {Benz}, {Megeath}, \& {Wilson}}]{Wang2013}
{Wang}, K.~S., {Bourke}, T.~L., {Hogerheijde}, M.~R., {et~al.} 2013, \aap, 558,
  A69

\bibitem[{{Weilbacher} {et~al.}(2015){Weilbacher}, {Monreal-Ibero},
  {Kollatschny}, {Ginsburg}, {McLeod}, {Kamann}, {Sandin}, {Palsa}, {Wisotzki},
  {Bacon}, {Selman}, {Brinchmann}, {Caruana}, {Kelz}, {Martinsson},
  {P{\'e}contal-Rousset}, {Richard}, \& {Wendt}}]{Weilbacher2015}
{Weilbacher}, P.~M., {Monreal-Ibero}, A., {Kollatschny}, W., {et~al.} 2015,
  \aap, 582, A114

\bibitem[{{Wells} {et~al.}(2015){Wells}, {Pel}, {Glasse}, {Wright},
  {Aitink-Kroes}, {Azzollini}, {Beard}, {Brandl}, {Gallie}, {Geers}, {Glauser},
  {Hastings}, {Henning}, {Jager}, {Justtanont}, {Kruizinga}, {Lahuis}, {Lee},
  {Martinez-Delgado}, {Mart{\'\i}nez-Galarza}, {Meijers}, {Morrison},
  {M{\"u}ller}, {Nakos}, {O'Sullivan}, {Oudenhuysen}, {Parr-Burman}, {Pauwels},
  {Rohloff}, {Schmalzl}, {Sykes}, {Thelen}, {van Dishoeck}, {Vandenbussche},
  {Venema}, {Visser}, {Waters}, \& {Wright}}]{Wells2015}
{Wells}, M., {Pel}, J.~W., {Glasse}, A., {et~al.} 2015, \pasp, 127, 646

\bibitem[{{White} \& {Sandell}(1995)}]{White1995}
{White}, G.~J. \& {Sandell}, G. 1995, \aap, 299, 179

\bibitem[{{Witt} {et~al.}(2006){Witt}, {Gordon}, {Vijh}, {Sell}, {Smith}, \&
  {Xie}}]{Witt2006}
{Witt}, A.~N., {Gordon}, K.~D., {Vijh}, U.~P., {et~al.} 2006, \apj, 636, 303

\bibitem[{{Yan} \& {Babb}(2023)}]{Yan2023}
{Yan}, P.-G. \& {Babb}, J.~F. 2023, \mnras, 522, 1265

\bibitem[{{Zannese} {et~al.}(2024){Zannese}, {Tabone}, {Habart}, {Goicoechea},
  {Zanchet}, {van Dishoeck}, {van Hemert}, {Black}, {Tielens}, {Veselinova},
  {Jambrina}, {Menendez}, {Verdasco}, {Aoiz}, {Gonzalez-Sanchez}, {Trahin},
  {Dartois}, {Bern{\'e}}, {Peeters}, {He}, {Sidhu}, {Chown}, {Schroetter}, {Van
  De Putte}, {Canin}, {Alarc{\'o}n}, {Abergel}, {Bergin}, {Bernard-Salas},
  {Boersma}, {Bron}, {Cami}, {Dicken}, {Elyajouri}, {Fuente}, {Gordon}, {Issa},
  {Joblin}, {Kannavou}, {Khan}, {Languignon}, {Le Gal}, {Maragkoudakis},
  {Meshaka}, {Okada}, {Onaka}, {Pasquini}, {Pound}, {Robberto}, {R{\"o}llig},
  {Schefter}, {Schirmer}, {Vicente}, \& {Wolfire}}]{Zannese2024}
{Zannese}, M., {Tabone}, B., {Habart}, E., {et~al.} 2024, Nature Astronomy
  [\eprint[arXiv]{2312.14056}]

\end{thebibliography}

\begin{appendix}
\section{Complementary tables and figure}

\begin{table*}
\caption{Field descriptions (the nomenclature and A$_{\rm V}$ values are taken from \citealt{Peeters2023}).}
\label{fields}
\begin{tabular}{lccccccc}\\
\hline
\noalign{\smallskip}
ID &  RA(J2000)  & Dec(J2000)  &  Size   &     PA     &  A$_{\rm V}$(bar)$^1$ &    A$_{\rm V}$(bar)$^2$ &    A$_{\rm V}$(foreground)$^3$   \\ \hline
%
H II region          &  05:35:20.187     &   $-$05:24:59.81       &  1.5$"$$\times$6$"$  &   40   &  0        &  0          & 1.5  \\
Atomic               &  05:35:20.259     &   $-$05:25:02.52       &   3$"$$\times$6$"$  &    40    &  4.33   &  7.86     & 1.3  \\
DF 1                   &  05:35:20.512     &   $-$05:25:11.95       &   3$"$$\times$6$"$  &    40    &  9.34   &  37.34   & 1.2 \\
DF 2                   &  05:35:20.615     &   $-$05:25:14.71       &  4$"$$\times$6$"$   &    40    &  4.67   &  8.67     & 1.2 \\
DF 3                   &  05:35:20.745     &   $-$05:25:20.56       &  4$"$$\times$6$"$   &    40    &  2.00   &  3.22     & 1.4 \\ \hline  
\end{tabular}

\noindent
$^1$Internal PDR extinction calculated using  the foreground formalism;  $^2$Internal PDR extinction calculated using  the intermingled formalism; $^3$ Foreground extinction. Nomenclature and A$_{\rm V}$ values are taken from \citet{Peeters2023}.
\end{table*}
 
\begin{table*}
\caption{Uncorrected line fluxes (erg cm$^{-2}$ s$^{-1}$ sr$^{-1}$ ).}
\label{fluxes}
\begin{tabular}{ccccc}\\
\hline
\noalign{\smallskip}
ID &   H$_2$ (17.035 $\mu$m)$^a$  & SI (25.24 $\mu$m)   & SIII (18.7 $\mu$m )$^a$ & SIV (10.5 $\mu$m)$^a$    \\
     &  $\times$ 10$^{-4}$
     &  $\times$ 10$^{-5}$  
     &  $\times$ 10$^{-2}$   
     &  $\times$ 10$^{-3}$   \\ \hline
HII region    &    2.00   &   $<$ 1.2$^b$   &  6.02   &  3.10 \\
Atomic        &    1.63    &   $<$ 2.4$^b$  &  4.07   &  4.18 \\
DF1            &    2.45    &    $<$ 2.3$^b$   &  1.80   &  2.75 \\
DF2            &    7.17    &    5.7 (0.9)  &  1.58   &  2.50 \\ 
DF3            &    7.81    &   11.4 (0.7) &  1.27   &  1.94 
\\ \hline 
\end{tabular}

\noindent
$^a$ The error is dominated by the calibration uncertainty (10\%).

\noindent
$^b$ 3$\times$rms upper limits.

\end{table*}

\begin{table*}
\caption{Extinction-corrected fluxes (erg cm$^{-2}$ s$^{-1}$ sr$^{-1}$ ).}
\label{corrs}
\begin{tabular}{ccccc}\\
\hline
\noalign{\smallskip}
ID &  H$_2$ (17.035 $\mu$m)$^a$ & [S I] (25.249 $\mu$m)$^a$   & [S III] (18.7 $\mu$m )$^b$ & [S IV] (10.5 $\mu$m)$^b$    \\
     &  $\times$ 10$^{-4}$
     &  $\times$ 10$^{-5}$  
     &  $\times$ 10$^{-2}$   
     &  $\times$ 10$^{-3}$   \\ \hline
HII region    &   2.09              &   $<$ 1.2           &  6.29    &  3.39 \\ 
Atomic        &   1.92$-$2.12   &   $<$ 2.8           &  4.23   & 4.51  \\
DF1            &    3.33$-$7.50  &   $<$ 3.7         &  1.86   &  2.95  \\
DF2            &    8.50$-$9.55  &  6.14$-$6.46     &  1.64   &  2.68   \\ 
DF3            &    8.62$-$8.93  &  11.90$-$12.09  &  1.32   &  2.11  
\\ \hline 
\end{tabular}

\noindent
$^a$ Lower and upper limits to the extinction-corrected fluxes obtained using [A$_{\rm V}$ (foreground) $+$ A$_{\rm V}$ (bar)$^1$] and [A$_{\rm V}$ (foreground) $+$ A$_{\rm V}$ (bar)$^2$], respectively. For the upper limits we only used  A$_{\rm V}$ (foreground) $+$ A$_{\rm V}$ (bar)$^2$. \\
$^b$ Extinction-corrected fluxes obtained using A$_{\rm V}$ (foreground).
\end{table*}

\begin{figure*}
\includegraphics[angle=0,scale=.8]{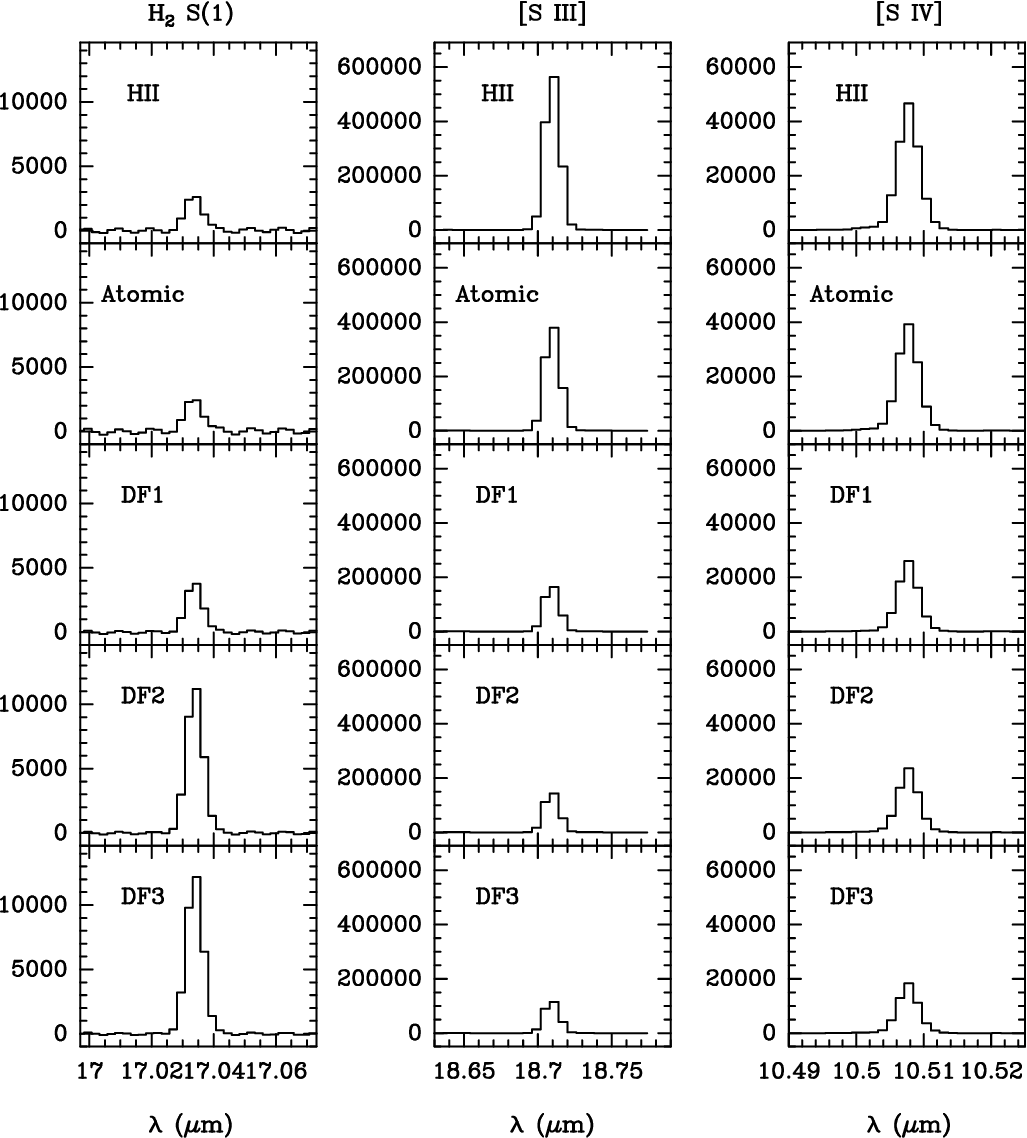}
\caption{Average intensity spectra of the H$_2$ S(1), [S III] 18.713 $\mu$m, and
[S IV] 10.5105 $\mu$m lines in the fields shown in Fig.~\ref{spectra}. The intensity scale is MJy sr$^{-1}$. }
\label{spectratodos}
\end{figure*}

\begin{table*}
\caption{Input parameters for chemical models.$^1$}
\label{mod}
\begin{tabular}{lll}\\
\hline \hline
Model   & Parameter  &   Value     \\
\hline 
1    &  G$_0$ at IF                                     &   3.1 $\times$ 10$^4$  \\         
      &   G$_0$ at the back side                &  0.1 \\                                  
      &  Inclination angle ($i$)                      &   60$^\circ$                  \\                                                                                           
      &  Thermal pressure (P$_{\rm th}$)    &    2.8 $\times$ 10$^7$  K cm$^{-3}$  \\                                      
      &  N$_H$/E(B-V)                                 &   1.05 $\times$ 10$^{22}$ cm$^{2}$ mag$^{-1}$   \\
      & $\zeta_{H_2}$                                  &   5 $\times$ 10$^{-17}$ s$^{-1}$                         \\
      & S/H                                                   &   1.86 $\times$ 10$^{-5}$    \\ \hline
2    &  G$_0$ at IF                                     &   3.1 $\times$ 10$^4$  \\         
      &   G$_0$ at the back side                 &  0.1 \\                                  
      &  Inclination angle ($i$)                      &   60$^\circ$                  \\                                                                                           
      &  Thermal pressure (P$_{\rm th}$)    &    2.8 $\times$ 10$^8$  K cm$^{-3}$  \\                                      
      &  N$_H$/E(B-V)                                 &   1.05 $\times$ 10$^{22}$ cm$^{2}$ mag$^{-1}$   \\
      & $\zeta_{H_2}$                                  &   5 $\times$ 10$^{-17}$ s$^{-1}$                         \\
      & S/H                                                   &   1.86 $\times$ 10$^{-5}$    \\ \hline
3     &  G$_0$ at IF                                     &   3.1 $\times$ 10$^4$  \\         
      &   G$_0$ at the back side.                 &  0.1 \\                                  
      &  Inclination angle ($i$)                      &   60$^\circ$                  \\                                                                                           
      &  Thermal pressure (P$_{\rm th}$)    &    2.8 $\times$ 10$^9$  K cm$^{-3}$  \\                                      
      &  N$_H$/E(B-V)                                 &   1.05 $\times$ 10$^{22}$ cm$^{2}$ mag$^{-1}$   \\
      & $\zeta_{H_2}$                                  &   5 $\times$ 10$^{-17}$ s$^{-1}$                         \\
      & S/H                                                   &   1.86 $\times$ 10$^{-5}$    \\ \hline
4    &  G$_0$ at IF                                     &   3.1 $\times$ 10$^3$  \\         
      &   G$_0$ at the back side                 &  0.1 \\                                  
      &  Inclination angle ($i$)                      &   60$^\circ$                  \\                                                                                           
      &  Thermal pressure (P$_{\rm th}$)    &    2.8 $\times$ 10$^8$  K cm$^{-3}$  \\                                      
      &  N$_H$/E(B-V)                                 &   1.05 $\times$ 10$^{22}$ cm$^{2}$ mag$^{-1}$   \\
      & $\zeta_{H_2}$                                  &   5 $\times$ 10$^{-17}$ s$^{-1}$                         \\
      & S/H                                                   &   1.86 $\times$ 10$^{-5}$    \\ \hline
5    &  G$_0$ at IF                                     &   3.1 $\times$ 10$^5$  \\         
      &   G$_0$ at the back side                 &  0.1 \\                                  
      &  Inclination angle ($i$)                      &   60$^\circ$                  \\                                                                                           
      &  Thermal pressure (P$_{\rm th}$)    &    2.8 $\times$ 10$^8$  K cm$^{-3}$  \\                                      
      &  N$_H$/E(B-V)                                 &   1.05 $\times$ 10$^{22}$ cm$^{2}$ mag$^{-1}$   \\
      & $\zeta_{H_2}$                                  &   5 $\times$ 10$^{-17}$ s$^{-1}$                         \\
      & S/H                                                   &   1.86 $\times$ 10$^{-5}$    \\ \hline
6    &  G$_0$ at IF                                     &   3.1 $\times$ 10$^4$  \\         
      &   G$_0$ at the back side                  &  0.1 \\                                  
      &  Inclination angle ($i$)                      &   60$^\circ$                  \\                                                                                           
      &  Thermal pressure (P$_{\rm th}$)    &    2.8 $\times$ 10$^8$  K cm$^{-3}$  \\                                      
      &  N$_H$/E(B-V)                                 &   1.05 $\times$ 10$^{22}$ cm$^{2}$ mag$^{-1}$   \\
      & $\zeta_{H_2}$                                  &   5 $\times$ 10$^{-16}$ s$^{-1}$                         \\
      & S/H                                                   &   1.86 $\times$ 10$^{-5}$    \\ \hline
7    &  G$_0$ at IF                                     &   3.1 $\times$ 10$^4$  \\         
      &   G$_0$ at the back side                  &  0.1 \\                                  
      &  Inclination angle ($i$)                      &   60$^\circ$                  \\                                                                                           
      &  Thermal pressure (P$_{\rm th}$)    &    2.8 $\times$ 10$^8$  K cm$^{-3}$  \\                                      
      &  N$_H$/E(B-V)                                 &   1.05 $\times$ 10$^{22}$ cm$^{2}$ mag$^{-1}$   \\
      & $\zeta_{H_2}$                                  &   5 $\times$ 10$^{-17}$ s$^{-1}$                         \\
      & S/H                                                   &   1.86 $\times$ 10$^{-7}$    \\ \hline
8    &  G$_0$ at IF                                     &   3.1 $\times$ 10$^4$  \\         
      &   G$_0$ at the back side                  &  0.1 \\                                  
      &  Inclination angle ($i$)                      &   60$^\circ$                  \\                                                                                           
      &  Thermal pressure (P$_{\rm th}$)    &    2.8 $\times$ 10$^8$  K cm$^{-3}$  \\                                      
      &  N$_H$/E(B-V)                                 &   1.6 $\times$ 10$^{22}$ cm$^{2}$ mag$^{-1}$   \\
      & $\zeta_{H_2}$                                  &   5 $\times$ 10$^{-17}$ s$^{-1}$                         \\
      & S/H                                                   &   1.86 $\times$ 10$^{-5}$    \\ 
 \hline \hline
\end{tabular}

\noindent
$^1$ The parameters not shown here are the same as in the reference model (see Table~\ref{refmod})
\end{table*}
\end{appendix}
\end{document}